# Using First-Order Logic to Reason about Policies

Joseph Y. Halpern and Vicky Weissman

Cornell University

A policy describes the conditions under which an action is permitted or forbidden. We show that a fragment of (multi-sorted) first-order logic can be used to represent and reason about policies. Because we use first-order logic, policies have a clear syntax and semantics. We show that further restricting the fragment results in a language that is still quite expressive yet is also tractable. More precisely, questions about entailment, such as 'May Alice access the file?', can be answered in time that is a low-order polynomial (indeed, almost linear in some cases), as can questions about the consistency of policy sets.

## 1. INTRODUCTION

A policy describes the conditions under which an action, such as reading a file, is permitted or forbidden. Digital-content providers have a rough idea of what their policies should be. Unfortunately, policies are typically described informally. As a result, their meaning and consequences are not always clear. To better understand the problem, consider the statement "only librarians may edit the on-line catalog". We can view this statement as a policy because it governs who may edit the catalog, based on whether the editor is a librarian. It is not clear if this policy permits librarians to make changes to the catalog or only forbids anyone who is not a librarian from doing so. The policy could be rewritten to remove this particular ambiguity, but others are likely to exist if policies are written in a natural language.

Of course, policies do not need to be written in a natural language. Access control lists (ACLs) [Pfleeger 1997] have been used for decades to capture simple policies in an unambiguous way. Unfortunately, ACLs lack the expressive power needed

Authors' address: J. Halpern and V. Weissman, Cornell University, Ithaca, NY 14853.

Authors supported in part by NSF under grant CTC-0208535, by ONR under grants N00014-00-1-03-41 and N00014-01-10-511, by the DoD Multidisciplinary University Research Initiative (MURI) program administered by the ONR under grant N00014-01-1-0795, and by AFOSR under grant F49620-02-1-0101.

A preliminary version of this paper appeared at the 16th IEEE Computer Security Foundations Workshop in Pacific Grove, California, 2003.

by many of today's digital-content providers. For example, we cannot capture the policy "members are permitted to access the digital library"; the best we can do using ACLs is to maintain a list that must be updated whenever the set of members change. Another option is to write policies in an XML-based language. Two popular choices are XrML (eXtensible rights Markup Language) (XrML) [ContentGuard 2001] and ODRL (Open Digital Rights Language) [Iannella 2001]. These languages can be given formal semantics (in part because their syntax, unlike that of natural languages, is quite restricted) and they do have more expressive power than ACLs. Prior to our work, however, neither language had formal semantics and, as a result, policies written in these languages were often ambiguous.[1]

The formal-methods community has proposed a number of languages that have formal semantics. Perhaps the most popular approach is to base the language on some extension of Datalog [Garcia-Molina et al. 2002]. These extensions are tractable fragments of first-order logic that allow a limited use of function symbols and negation. Unfortunately, the extensions do not seem to have the necessary expressive power to capture a number of policies that are currently written in English. For example, in the iTunes Terms of Sale [Apple Computer 2004], certain actions are explicitly forbidden and others are unregulated; most of the variants of Datalog cannot distinguish between the two categories.

Our goal in this paper is to provide a better language in which to write policies. The language must have a clear syntax and semantics. To be of practical interest, it must also satisfy (at least) the following desiderata.

(1) It must be expressive enough to capture in an easy and natural way the policies that people want to discuss.

(2) It must be tractable enough to allow interesting queries about policies to be answered efficiently.

(3) It must be usable by non-experts, because we cannot expect policymakers and administrators to be well-versed in logic or programming languages.

To achieve our objectives, we use a fragment of first-order logic that we call *Lithium*. While our approach is reminiscent of the Datalog ones, the restrictions that we make are quite different from those made previously. We believe (and will argue throughout this paper) that the resulting language is especially well-suited for many applications, and has a number of advantages over variants of Datalog.

Because Lithium is a fragment of first-order logic, it automatically has a clear syntax and semantics; thus, it remains to argue that the logic satisfies the three goals listed above. Whether a logic is sufficiently expressive to satisfy the first objective naturally depends on the application. To evaluate our approach, we gathered a large collection of policies from different types of libraries, ranging from small public libraries to the Library of Congress; we then determined that the policies could be written in Lithium (sometimes considering multiple versions of the policies, one for each interpretation). In addition, we have performed the same analysis on various parts of US legislation, including substantial fragments of the Privacy Rule, which governs access to electronic medical files, and Title 42, Chapter 7 of the US Code, which determines who is eligible for Social Security. Finally, we have shown that large fragments of XrML and ODRL can be translated into Lithium; see [Halpern and Weissman 2004; Pucella and Weissman 2004]. Our work indicates that Lithium is expressive enough to capture many (if not most) policies of interest.

---

[1]Recently, we have given formal semantics to a representative fragment of XrML [Halpern and Weissman 2004] and a representative fragment of ODRL [Pucella and Weissman 2004]. Both semantics are based on the approach given in this paper. The latest specification of XrML [MPEG 2004] includes formal semantics, in part to address our criticisms, but their approach is different from that used here.

For the second desideratum, we focus on two key queries:

—Given a set of policies and an *environment* that provides all relevant facts (e.g., "Alice is a librarian", "Anyone who is a librarian for less than a year is a novice", etc.), does it follow that a particular action, such as Alice editing the on-line catalog, is permitted or forbidden?

—Is a set of policies consistent? In other words, are there no actions that are both permitted and forbidden by the policies in the set? This question is particularly interesting for collaboration. For example, suppose that Alice is writing the policies for her university's new outreach program. If the union of her policies and the university policies is consistent, then she knows that her policies do not contradict those of the university.

The answers to these questions could be used by enforcement mechanisms and individuals who want to do regulated activities. More importantly, we believe that the answers provide a reasonably good understanding of the policies, increasing our confidence that the formal statements capture the informal rules and the informal rules capture the policy creator's intent.

The rest of this paper is organized as follows. In the next section, we formally define our notions of policy and environment. We also give examples that illustrate how policies can be represented in an appropriate fragment of first-order logic. Sections 3 and 4 focus on queries about permissions (that is, whether a specific action is permitted given an environment and a set of policies); all our results hold with essentially no change for queries about prohibitions. We show in Section 3 that such queries are, in general, hard to answer. In Section 4, we consider some restrictions that we believe typically hold in practice; under these restrictions, the queries are tractable. We address the consistency problem in Section 5. Of course, we do not expect typical users to be experts in first-order logic. In Section 6, we discuss what can be done to make Lithium accessible to non-logicians. Related work, including the Datalog approaches, are discussed in Section 7. Our concluding remarks are given in Section 8. Most proofs are left to the appendix.

## 2. A FIRST-ORDER LOGIC FOR REASONING ABOUT POLICIES

For the rest of the paper, we assume knowledge of first-order logic at the level of Enderton [1972]. More specifically, we assume the reader is familiar with the syntax of first-order logic, including constants, variables, predicate symbols, function symbols, and quantification; with the semantics of first-order logic, including relational models and valuations; and with the notions of satisfiability and validity of first-order formulas. Recall that many-sorted first-order logic is first-order logic modified so that each term is associated with a sort (i.e., type); variables of sort $s$ range over the elements of sort $s$; and the signatures of predicate and function symbols restrict each argument to elements of a particular sort.

We use many-sorted first-order logic with equality over some vocabulary $\Phi$ to express and reason about policies. Let $\mathcal{L}^{fo}(\Phi)$ denote the set of first-order formulas over the vocabulary $\Phi$. For this paper, we assume that there are at least three sorts, *Actions* (e.g., accessing a file), *Subjects* (the agents that perform actions; these are sometimes called *principals* in the literature), and *Times*. While these sorts seem natural for any policy logic, other sorts may be desired for particular applications. These sorts, including objects and roles, may be added to the logic without affecting our results.

The vocabulary $\Phi$ is application-dependent; however, we assume that $\Phi$ contains a constant **now** of sort *Times* and a binary predicate **Permitted** on *Subjects* × *Actions*. The constant **now** denotes the current time. In practice, a global clock would determine the interpretation of **now**. **Permitted**$(t, t')$ means that subject $t$ is allowed to perform action $t'$. In practice, it may be useful to add additional arguments to **Permitted**, such as when the action is permitted and who is authorizing the granting or revoking of the permission. Which arguments are added (if any) depend on the application and on other choices made in the vocabulary. For example, consider a policy $p$ that says "Alice is permitted to append image $m$ to file $f$". We could either take **Permitted** to be a binary predicate and **append** to be a binary function, and express $p$ as "**Permitted**(**Alice**, **append**$(m, f)$)"; or we could take **append** to be a unary function and **Permitted** to be a ternary predicate, and express $p$ as "**Permitted**(**Alice**, **append**$(m), f$)". Our results apply regardless of which choice is made, because they do not depend on the arity of **Permitted** and the other functions and predicates in the language. In fact, our results still hold even if policies refer to different variants of **Permitted**, with different arities.

A *policy* is a closed first-order formula of the form

$$\forall x_1 \ldots \forall x_m (f \Rightarrow (\neg)\mathbf{Permitted}(t, t')),$$

where $f$ is any first-order formula, $t$ and $t'$ are terms of sort Subject and Action respectively, and the notation $(\neg)\mathbf{Permitted}$ indicates that the **Permitted** predicate may or may not be negated. Defining a policy in this way provides a structure that matches our intuition, namely, that a policy is a set of conditions under which an action is or is not permitted.

To illustrate how policies can be expressed in first-order logic, consider the following examples.

EXAMPLE 2.1. *The policy "only librarians may edit the catalog" can be characterized by the following two formulas:*

$$\forall x(\neg\mathbf{Librarian}(x) \Rightarrow \neg\mathbf{Permitted}(x, \mathbf{edit\ the\ catalog}))$$
$$\forall x(\mathbf{Librarian}(x) \Rightarrow \mathbf{Permitted}(x, \mathbf{edit\ the\ catalog})).$$

*(Depending on the intended meaning of the English statement, the first formula by itself may characterize the policy.)* ▮

EXAMPLE 2.2. *The policy "a customer may download any article if she has paid a fee within the past six weeks" can be rewritten as "if an individual $i$ has paid the fee within the past six weeks, $i$ is a customer, and $a$ is some article, then $i$ may* *download a". The policy can be encoded readily as*

$\forall i \forall t \forall a((\textbf{PaidFee}(i,t) \wedge (\textbf{\textit{now}} - 6 < t < \textbf{\textit{now}}) \wedge \textbf{Customer}(i, \textbf{\textit{now}}) \wedge \textbf{Article}(a)$
$\Rightarrow \textbf{Permitted}(i, \textbf{download}(a))).$ ∎

EXAMPLE 2.3. *The policy set "anyone may sing" and "anyone who is allowed to sing may dance" can be characterized by the following two formulas:*

$$\forall x(\textbf{Permitted}(x, \textbf{sing})$$
$$\forall x(\textbf{Permitted}(x, \textbf{sing}) \Rightarrow \textbf{Permitted}(x, \textbf{dance})).$$ ∎

To determine the consequences of policies, we need to know which facts are true in the *environment* (i.e., the context in which the policies are applied). For example, if the environment implies that Alice is a librarian, then the policies in Example 2.1 imply that she may edit the catalog. If the environment is silent as to whether Alice is a librarian, then the policies in Example 2.1 do not regulate her actions. The *environment* may include specific statements such as "Alice is a librarian", "*The Cat in the Hat* is a children's book", or "Sally has a junior library card". General statements may also be included, such as the conditions under which a customer is considered to be in good standing and "at all times, there is a senior staff member who is on call". All the examples we have considered so far confirm our belief that first-order logic is sufficiently expressive to capture most environments that are likely to arise in practice. Thus, we formally define an environment to be a closed first-order formula that does not contain the **Permitted** predicate. The requirement that the environment not contain **Permitted** encourages the intuitive separation between the environment, which is a description of reality, and the policies, which are the rules governing that reality.

The two types of queries discussed in the introduction can now be formalized. The first query, is an individual $t$ permitted to perform an action $t'$ (where $t$ and $t'$ are closed terms) given an environment $E$ and some policies $p_1, \ldots, p_n$, amounts to asking if the formula $E \wedge p_1 \wedge \ldots \wedge p_n \Rightarrow \textbf{Permitted}(t, t')$ is valid. (Similarly, $t$ is forbidden to do $t'$ if and only if $E \wedge p_1 \wedge \ldots \wedge p_n \Rightarrow \neg\textbf{Permitted}(t, t')$ is valid.) The second query, "Are the policies consistent?", asks if the formula $E \wedge p_1 \wedge \ldots \wedge p_n$ is satisfiable. For ease of exposition, we focus on determining if an action is permitted (or forbidden). As we show, it is easy to modify our techniques to handle the consistency question.

## 3. INTRACTABILITY RESULTS

In general, the queries in which we are interested cannot be answered efficiently. Indeed, the problem in its full generality is easily seen to be undecidable if the vocabulary $\Phi$ has at least one binary predicate other than **Permitted** (and closed terms $t$ and $t'$ of sort *Subjects* and *Actions*, respectively, so that it is possible to actually form queries). To see this, let $f$ be an arbitrary formula that does not contain **Permitted**. Consider the policy $f \Rightarrow \textbf{Permitted}(t, t')$, and let the environment be empty (i.e., **true**). Standard manipulations show that

$$(f \Rightarrow \textbf{Permitted}(t, t')) \Rightarrow \textbf{Permitted}(t, t')$$

is equivalent to

$$f \vee \textbf{Permitted}(t, t').$$

Since $f$ does not mention **Permitted**, the last formula is valid iff $f$ is valid. The validity problem for first-order formulas is well known to be undecidable, even if we restrict to formulas whose only nonlogical symbol is a binary predicate. In fact, undecidability holds if we further restrict to formulas that have a single alternation of quantifiers (i.e., formulas of the form $Q_1 x_1 \ldots Q_n x_n R_1 y_1 \ldots R_m y_m f$, where $Q_i = \exists$ and $R_j = \forall$ for $i = 1, \ldots, n$ and $j = 1, \ldots, m$ or vice-versa, and $f$ is quantifier-free) [Börger et al. 1997]. So, in general, we cannot determine whether a single policy implies a permission if writing the policy as a first-order formula requires an alternation of quantifiers and a binary predicate other than **Permitted**. In fact, undecidability holds even without the assumption that $\Phi$ has a binary predicate other than **Permitted**.

THEOREM 3.1. *Let $\mathcal{L}_0$ be the set of closed formulas of the form*

$$(f \Rightarrow \textbf{Permitted}(c, c')) \Rightarrow \textbf{Permitted}(c, c'),$$

*where $c$ and $c'$ are constants of the appropriate sorts, $f$ has a single alternation of quantifiers, and the only nonlogical symbol in $f$ is* **Permitted***. The validity question for $\mathcal{L}_0$ is undecidable.*

Not surprisingly, similar undecidability results hold if we allow formulas in the environment to involve an alternation of quantifiers (provided that there is a binary predicate in the language other than **Permitted**, since we do not allow **Permitted** in the environment). Given Theorem 3.1, it seems that our only hope is to forbid any alternation of quantifiers.

How much quantification do we really need? A quantifier-free environment suffices to capture simple databases. Many applications, however, need a richer environment that includes general properties, such as "men are not women" and "a senior citizen is anyone over sixty-five years old". For these applications, universal quantification is needed in the environment. In addition, almost all applications need quantification in their policies. To see why, notice that if we do not allow a policy to have any quantification (i.e., define a policy to have the form $f \Rightarrow \textbf{Permitted}(t, t')$ where $t$ and $t'$ are closed terms and $f$ is quantifier-free), then each policy must govern a specific individual and action. For example, we can say "If Alice is good, she may play outside", but we cannot say "All good children may play outside". Because policies typically permit an individual to do an action based on the attributes of that individual, we must allow policies to be universally quantified.

All policies expressible in XrML and in ODRL, as well as the policies that we have collected from libraries and government databases, can be written as universal formulas (i.e., as formulas that can be written in the form $\forall x_1 \ldots \forall x_n f$, where $f$ is quantifier-free). Some of the policies that we collected may appear to need existential quantification, but they can be converted to equivalent universal formulas. Example 3.2 illustrates how we can apply standard first-order transformations to do the conversion.

EXAMPLE 3.2. *Consider the policy "anyone who is accompanied by a librarian may enter the stacks". A natural way to state this in first-order logic is*

$$\forall x_1 (\exists x_2 (\textbf{Librarian}(x_2) \wedge \textbf{Accompanies}(x_2, x_1)) \Rightarrow \textbf{Permitted}(x_1, \textbf{enter}(\textbf{stacks}))).$$

*This formula is logically equivalent to*

$$\forall x_1 \forall x_2((\textbf{Librarian}(x_2) \wedge \textbf{Accompanies}(x_2, x_1)) \Rightarrow \textbf{Permitted}(x_1, \textbf{enter}(\textbf{stacks}))),$$

*which uses only universal quantification.* ∎

Note that **enter** is a function in Example 3.2. Unfortunately, it is well known that the validity problem for existential formulas with function symbols is undecidable, even if we restrict to formulas with only two existentials and one unary function symbol [Börger et al. 1997]. The following strengthening of Theorem 3.1 is almost immediate.

THEOREM 3.3. *Let $\mathcal{L}_1$ be the set of closed formulas of the form*

$$\forall x_1 \forall x_2(f \Rightarrow \textbf{Permitted}(c, c')) \Rightarrow \textbf{Permitted}(c, c'),$$

*where $c$ and $c'$ are constants of the appropriate sort and $f$ is a quantifier-free formula whose only nonlogical symbols are **Permitted** and a unary function. The validity problem for $\mathcal{L}_1$ is undecidable.*

Theorem 3.3 suggests that even if we drastically reduce quantification, we still need to disallow functions to get decidability. Once we severely restrict quantification and remove functions entirely, then we do get a decidable fragment, but it is not tractable. Recall that $\Pi_2^P$ is the second level of the polynomial hierarchy, and represents languages that can be decided in co-NP with an NP oracle.

THEOREM 3.4. *Let $\Phi$ be a vocabulary that contains **Permitted**, constants $c$ and $c'$ of sorts Subjects and Actions, respectively, and possibly other predicate and constant symbols (but no function symbols). Assume that there is a bound on the arity of the predicate symbols in $\Phi$ (that is, there exists some $N$ such that all predicate symbols in $\Phi$ have arity at most $N$). Finally, let $\mathcal{L}_2$ be the set of all closed formulas in $\mathcal{L}^{fo}(\Phi)$ of the form $E \wedge p_1 \wedge \ldots \wedge p_n \Rightarrow \textbf{Permitted}(c, c')$ such that $E$ is a conjunction of quantifier-free and universal formulas and each policy $p_1, \ldots, p_n$ has the form $\forall x_1 \ldots \forall x_m(f \Rightarrow \textbf{Permitted}(t_1, t_2))$, where $t_1$ and $t_2$ are terms of the appropriate sort and $f$ is quantifier-free.*

(a) *The validity problem for $\mathcal{L}_2$ is in $\Pi_2^P$.*

(b) *If $\mathcal{L}_3$ is the set of formulas in $\mathcal{L}_2$ in which every policy's antecedent is a conjunction of literals, then the validity problem for $\mathcal{L}_3$ is $\Pi_2^P$ hard.*

(c) *If $\mathcal{L}_4$ is the set of $\mathcal{L}_2$ formulas in which $E$ is quantifier-free, then the validity problem for $\mathcal{L}_4$ is both NP-hard and co-NP hard.*

We remark that if we do not require the arity of the predicate symbols in $\Phi$ to be bounded, then we must replace $\Pi_2^P$ by co-NEXPTIME (co-nondeterministic exponential time) in parts (a) and (b) [Börger et al. 1997].

Theorems 3.1, 3.3, and 3.4 seem to suggest that the questions we are interested in are hopelessly intractable. Fortunately, things are not nearly as bad as they seem.

## 4. IDENTIFYING TRACTABLE SUBLANGUAGES

The work on Datalog and its variants mentioned in the introduction demonstrates that there are useful, tractable fragments of first-order logic. In this section we define Lithium, a fragment of first-order logic characterized by a different set of restrictions than those considered by the Datalog community, show that these restrictions lead to tractability, and argue that they are particularly well-suited to reasoning about policies.

As a first step towards defining Lithium, we characterize the classes of environments and policies that are likely to occur in practice. A *basic environment* is an environment that is a conjunction of ground literals. Basic environments are sufficiently expressive to capture the information in databases and certificates. While this is adequate for many applications, basic environments cannot represent general properties such as "every citizen of Germany is a member of the European Union". To capture these, we define a *standard environment* to be an environment that is a conjunction of ground literals and closed formulas of the form $\forall x_1 \ldots \forall x_n (\ell_1 \wedge \ldots \wedge \ell_k \Rightarrow \ell_{k+1})$, where $\ell_1, \ldots, \ell_{k+1}$ are literals. Each conjunct of a standard environment is an *environment fact*. Note that every basic environment is a standard environment. A *standard policy* is a policy of the form $\forall x_1 \ldots \forall x_n (\ell_1 \wedge \ldots \wedge \ell_k \Rightarrow \mathbf{Permitted}(t_1, t_2))$, where $\ell_1, \ldots, \ell_{k+1}$ are literals and both $t_1$ and $t_2$ are terms of the appropriate sort. Standard environments and standard policies are sufficiently expressive for all of the applications that we have considered. A *simple policy* is a standard policy where none of the literals in the antecedent mentions $\mathbf{Permitted}$. For example, $\mathbf{Permitted}(t_1, t_2) \Rightarrow \mathbf{Permitted}(t_1, t_3)$ is not a simple policy.

A *policy base* is a formula of the form $E \wedge P$, where $E = E_0 \wedge E_1$ is a standard environment, $E_0$ is a conjunction of ground literals, $E_1$ is a conjunction of universally quantified formulas, and $P$ is a conjunction of standard policies. In the rest of the paper, when we write standard queries, we assume that the formulas $E$, $E_0$, $E_1$, and $P$ satisfy these constraints (that is, $E$ is a standard environment of the form $E_0 \wedge E_1$; $E_0$ is a basic environment; and so on). We are interested in characterizing policy bases $E \wedge P$ for which it is tractable to determine whether the query $E \wedge P \Rightarrow \mathbf{Permitted}(t, t')$ is valid, where $t$ and $t'$ are terms of the appropriate sort. We call such a query a *standard query*.

In the next section, we define a set of restrictions on standard queries that guarantee that validity can be determined quickly. After presenting the restrictions, we evaluate the likelihood that the restrictions will hold in practice. In subsequent sections, we relax each of the restrictions to accommodate a wider range of applications without sacrificing tractability. Roughly speaking, Lithium, which is formally defined in Section 4.2.5, is the set of standard queries that satisfy the relaxed restrictions.

## 4.1   A Tractable Sublanguage

We use the following terms to define the initial set of restrictions. A variable $v$ is *constrained* in a clause $c$ if $v$ appears as an argument to $\mathbf{Permitted}$ in $c$. For example, both $x$ and $y$ are constrained in the clause $\forall x \forall y \forall z (\neg \mathbf{R}(x, z) \vee \mathbf{Permitted}(x, y))$; $z$ is not constrained. Two literals $\ell$ and $\ell'$ are *unifiable* if there are variable substitutions $\sigma$ and $\sigma'$ such that $\ell\sigma = \ell'\sigma'$. For example, $\mathbf{R}(x, \mathbf{c_1})$ and $\mathbf{R}(\mathbf{c_2}, y)$ are unifiable by substituting $\mathbf{c_2}$ for $x$ and $\mathbf{c_1}$ for $y$, while $\mathbf{R}(x, \mathbf{c_1})$ and $\mathbf{R}(y, \mathbf{c_2})$ are not unifiable (if $c_1$ and $c_2$ are distinct constants). Let $f$ be a formula in CNF[2] and let $\ell$ be a literal in $f$. We say that $\ell$ is *bipolar* in $f$ if there is another literal $\ell'$ in $f$ such that $\ell$ and $\neg\ell'$ are unifiable. The pair $\ell, \ell'$ is called a *bipolar pair*. For example, consider the formula $f = \forall x(\mathbf{Permitted}(x, \mathbf{nap}) \Rightarrow \mathbf{Permitted}(\mathbf{Advisor}(x), \mathbf{nap}))$, which in CNF is $\forall x(\neg\mathbf{Permitted}(x, \mathbf{nap}) \vee \mathbf{Permitted}(\mathbf{Advisor}(x), \mathbf{nap}))$. Because $\neg\mathbf{Permitted}(x, \mathbf{nap})[x/\mathbf{Advisor}(y)] = \neg\mathbf{Permitted}(\mathbf{Advisor}(x), \mathbf{nap})[x/y]$, the literals $\neg\mathbf{Permitted}(x, \mathbf{nap})$ and $\mathbf{Permitted}(\mathbf{Advisor}(x), \mathbf{nap})$ are bipolar in $f$; together they form a bipolar pair.

THEOREM 4.1. *Let $\mathcal{L}_5$ consist of all standard queries of the form $E \wedge P \Rightarrow$ $\mathbf{Permitted}(t, t')$ such that*

*(1) $E$ is basic (i.e., $E$ is a conjunction of ground literals),*

*(2) there are no bipolar literals in $P$,*

*(3) equality is not mentioned in $E \wedge P$, and*

*(4) every variable appearing in a conjunct $p$ of $P$ is constrained in $p$.*

*The validity of formulas in $\mathcal{L}_5$ can be determined in time $O((|E|+|P||\mathbf{Permitted}(t, t')|)\log|E|)$, where $|\varphi|$ denotes the length of $\varphi$, when viewed as a string of symbols.*

Note that the language $\mathcal{L}_5$ includes formulas such as

$\mathbf{Student}(\mathbf{Alice}) \wedge \mathbf{Good}(\mathbf{Alice}) \wedge$
$\forall x(\mathbf{Student}(x) \Rightarrow \mathbf{Permitted}(x, \mathbf{work})) \wedge$
$\forall x(\mathbf{Student}(x) \wedge \mathbf{Good}(x) \Rightarrow \mathbf{Permitted}(x, \mathbf{play})) \Rightarrow \mathbf{Permitted}(\mathbf{Alice}, \mathbf{play})$

(may Alice play given that Alice is a student, Alice is good, all students may work, and all good students may play). Unlike Theorem 3.4(c), function symbols are allowed in Theorem 4.1. Moreover, there is no assumption that the arity of predicates and functions in the vocabulary is bounded. The price we pay for this added generality and for cutting the complexity to linear in the number of policies (which could well be large), linear in the length of the permission being considered (which is almost certainly small), and not much more than linear in the size of the database (which we expect to be relatively small, particularly in certificate-passing systems) is the four restrictions. We now discuss the likelihood that the restrictions will hold in practice; in subsequent sections we consider how the restrictions can be relaxed.

As we have already said, basic environments are sufficiently expressive to capture the facts stored in databases and certificates. They are also sufficiently expressive for the library applications that we considered and for the policies that can be written in XrML or ODRL, since both languages assume a minimal environment containing facts such as the current time, the time of the most recent revocation polling, and the number of times that a particular subject has done a specific action (e.g., printing a file). It is true, however, that basic environments are not always enough. For example, the documents that describe who may collect Social Security define an aged person to be anyone 65 years old or older, who is a resident of the U.S., and is either a citizen or an alien residing in the U.S. both legally and permanently. A basic environment cannot capture this definition.

---

[2]We say that a first-order formula is in CNF if it has the form $c_1 \wedge \ldots \wedge c_n$, where each $c_i$ has the form $Q_1 x_1 \ldots Q_m x_m(\varphi)$, each $Q_j \in \{\forall, \exists\}$, and $\varphi$ is a (quantifier-free) disjunction of literals, for $i = 1, \ldots, n$ and $j = 1, \ldots, m$. Each $\varphi$ is called a *clause*. We sometimes identify a universal formula in CNF with its set of clauses.

The second restriction, that there are no bipolar literals in $P$, is likely to hold if all the policies are *permitting policies* (that is, their conclusions have the form **Permitted**$(t_1, t_2)$) or all are *denying policies* (that is, their conclusions have the form ¬**Permitted**$(t_1, t_2)$). To see why, recall that a permitting policy says 'if the following conditions hold, then a particular action is permitted'. These conditions typically include requirements that someone possess one or more credentials, such as a library card or a driver's license. It is fairly rare that *not* having a credential, such as not having a driver's license, increases an individual's rights. Therefore, we do not expect credentials to correspond to bipolars. Similar arguments may be made for other types of information.

If the policy set includes a mix of permitting and denying policies, then it seems less likely that the bipolar restriction will hold. For example, suppose that an individual may smoke if and only if she is over eighteen years old. We could write this statement as two policies

$$p_1 = \forall x(\textbf{GreaterThan}(\textbf{age}(x), 18) \Rightarrow \textbf{Permitted}(x, \textbf{smoke})) \text{ and}$$
$$p_2 = \forall x(\neg\textbf{GreaterThan}(\textbf{age}(x), 18) \Rightarrow \neg\textbf{Permitted}(x, \textbf{smoke})).$$

Note that $p_1$ is a permitting policy, $p_2$ is a denying policy, and every literal in $p_1 \wedge p_2$ is bipolar in $p_1 \wedge p_2$.

The third restriction, that equality is not used, is satisfied by most of the policies and environment facts that we collected. However, the restriction is violated by threshold policies (e.g., "if Alice is blackballed by at least two people, then she may not join the club") and by statements that say two distinct names refer to the same individual (e.g., "**Alice Smith = wifeOf(Bob Smith)**" and "**number of accesses = 7**").

The last restriction, that every variable appearing in a policy $p$ is constrained in $p$, holds if an individual is granted or denied permission based solely on her attributes and the attributes of the regulated action. Notice that the policies in Examples 2.1 and 2.3 have this form, but the policies in Examples 2.2 and 3.2 do not. In particular, whether the policy in Example 3.2 allows $x_1$ to enter the stacks depends on an attribute of some other individual $x_2$.

Before relaxing the restrictions, we briefly discuss why they are sufficient for tractability. The first three restrictions allow us to consider each policy individually, that is, $E \wedge P \Rightarrow \textbf{Permitted}(t, t')$ is valid iff $E \wedge p \Rightarrow \textbf{Permitted}(t, t')$ is valid for some conjunct $p$ in $P$.

PROPOSITION 4.2. *Suppose that $E \wedge P \Rightarrow \textbf{Permitted}(t, t')$ is a standard query in which $E$ is basic, the equality symbol is not mentioned in $E \wedge P$, and there are no bipolars in $P$. Then $E \wedge P \Rightarrow \textbf{Permitted}(t, t')$ is valid iff there is a conjunct $p$ of $P$ such that $E \wedge p \Rightarrow \textbf{Permitted}(t, t')$ is valid.*

If the last restriction holds, then we can determine quickly whether $E \wedge p \Rightarrow$ **Permitted**$(t, t')$ is valid for some conjunct $p$ of $P$.

## 4.2   Relaxing the Restrictions

In this section, we consider the extent to which we can relax the four restrictions given in Theorem 4.1, while still maintaining tractability. We consider each of the restrictions in turn.

4.2.1   *Beyond Basic Environments.*  There is an obvious generalization of Theorem 4.1: we simply remove the first restriction and replace every reference to $P$ with $E_1 \wedge P$, where $E_1$ is the conjunction of universal statements in $E$. This results in three restrictions: there are no bipolar literals in $E_1 \wedge P$, equality is not mentioned in $E_1 \wedge P$, and every variable appearing in a conjunct $c$ of $E_1 \wedge P$ is constrained in $c$. Unfortunately, because **Permitted** does not appear in the environment, the variable restriction holds only if the environment has no quantification. In addition, we can prove that, if there are no bipolar literals in $E_1 \wedge P$, then $E \wedge P \Rightarrow$ **Permitted**$(t, t')$ is valid if and only if $E$ is inconsistent or $E_0 \wedge P \Rightarrow$ **Permitted**$(t, t')$ is valid, where $E_0$ is the conjunction of ground literals in $E$. This means that a universal statement in the environment can affect the validity of a query only if it makes the environment inconsistent. To support interesting universal statements in the environment, we must relax the restrictions on bipolar literals and variables, which we do in Sections 4.2.2 and 4.2.4, respectively.

4.2.2   *Relaxing the Bipolar Restriction.*  If we allow bipolar literals in $E_1 \wedge P$, then a permission might follow from a set of policies without following from any single policy. In other words, the conclusion of Proposition 4.2 might not hold.

Example 4.3.  *Consider two policies $p_1$ and $p_2$, where $p_1$ says "Alice may cry if she is happy" and $p_2$ says "Alice may cry if she is not happy". Formally,*

$$p_1 = \textbf{Happy}(\textbf{Alice}) \Rightarrow \textbf{Permitted}(\textbf{Alice}, \textbf{cry}) \ and$$
$$p_2 = \neg\textbf{Happy}(\textbf{Alice}) \Rightarrow \textbf{Permitted}(\textbf{Alice}, \textbf{cry}).$$

*Clearly, $p_1 \Rightarrow$ **Permitted**$(\textbf{Alice}, \textbf{cry})$ is not valid, because Alice might not be happy. Similarly, $p_2 \Rightarrow$ **Permitted**$(\textbf{Alice}, \textbf{cry})$ is not valid, because Alice might be happy. But $p_1 \wedge p_2 \Rightarrow$ **Permitted**$(\textbf{Alice}, \textbf{cry})$ is valid, because Alice is either happy, in which case she may cry by $p_1$, or she is not happy, in which case she may cry by $p_2$. So Alice's right to cry doesn't follow from either policy individually, but follows from both policies together, essentially because $p_1 \wedge p_2$ includes the bipolar pair $(\textbf{Happy}(\textbf{Alice}), \neg\textbf{Happy}(\textbf{Alice}))$.* ∎

Example 4.3 shows how we can use bipolar literals to infer a statement, namely Alice may cry, from two clauses, namely $p_1$ and $p_2$.  *Resolution* [Nerode and Shore 1997] generalizes the reasoning in this example.  To understand how resolution works, let $c$ be the clause $\forall x_1 \ldots \forall x_n (\ell \Rightarrow d)$ and let $c'$ be the clause $\forall x'_1 \ldots \forall x'_m (\ell' \Rightarrow d')$, where $\ell$ and $\ell'$ are literals. Suppose that $\sigma$ and $\sigma'$ are variable substitutions such that $\ell\sigma = \neg\ell'\sigma'$. It is easy to see that $c \wedge c' \Rightarrow d\sigma \vee d'\sigma'$ is valid. Using standard terminology, we call $c$ and $c'$ the *parents* of the *resolvent* $d\sigma \vee d'\sigma'$, and we say that $c$ and $c'$ *resolve on* $\ell\sigma$ to create $d\sigma \vee d'\sigma'$.[3]  The closure under resolution of a universal formula $f$, denoted $R(f)$, is the smallest set of clauses that includes the clauses in $f$ (when $f$ is in CNF) and is closed under resolution, that is if $e$ is the resolvent of two distinct clauses in $R(f)$, then $e$ is in $R(f)$. Roughly speaking, the resolvents in $R(f)$ are all the clauses that can be inferred from the clauses in $f$.

---

[3]Actually, the resolvent is created using a particular substitution, called a *most general unifier*, which is essentially the substitution that replaces variables with constants only when necessary. For example, the most general unifier for $c = \forall y(\neg\textbf{R}(y) \Rightarrow \textbf{S}(y))$ and $c' = \forall x(\textbf{R}(\textbf{f}(x)) \Rightarrow \textbf{S}(\textbf{g}(x)))$

Our interest in resolution is motivated in part because we can prove that a standard query $q$ of the form $E_0 \wedge E_1 \wedge P \Rightarrow \mathbf{Permitted}(t, t')$ that does mention equality is valid iff there is a clause $c \in R(E_1 \wedge P)$ such that $E_0 \wedge c \Rightarrow \mathbf{Permitted}(t, t')$ is valid. The role of the bipolar restriction in the language $\mathcal{L}_5$ is also best understood in the context of resolution. Part of our approach to guaranteeing tractability involves keeping $R(E_1 \wedge P)$ small. If there are no bipolar literals in $E_1 \wedge P$, then $R(E_1 \wedge P)$ includes only the conjuncts of $E_1$ and $P$; there are no resolvents. We can also prove that $R(E_1 \wedge P)$ is still fairly small if each conjunct in $E \wedge P$ has at most one bipolar literal. As a result, we maintain tractability if there is at most one bipolar literal in each conjunct (see Theorem 4.7). However, if even a single conjunct of $E_1 \wedge P$ has two bipolars, and the other conjuncts have at most one bipolar each, then $R(E_1 \wedge P)$ can be infinite.

EXAMPLE 4.4. *Suppose we have two policies; the first is "Alice may play" and the second is "for all individuals $x_1$ and $x_2$, if $x_1$ may play and $x_2$ is $x_1$'s boss, then $x_2$ may play". We can write these policies as*

$p_1 = \mathbf{Permitted}(\mathbf{Alice}, \mathbf{play})$
$p_2 = \forall x_1 \forall x_2 (\mathbf{Permitted}(x_1, \mathbf{play}) \wedge \mathbf{BossOf}(x_2, x_1) \Rightarrow \mathbf{Permitted}(x_2, \mathbf{play}))$

*It is not hard to see that for any integer $n$, the closure of $p_1 \wedge p_2$ includes the clause*

$$(\bigvee_{i=1,\dots,n} \neg\mathbf{BossOf}(x_i, x_{i-1})) \vee \neg\mathbf{BossOf}(x_0, \mathbf{Alice}) \vee \mathbf{Permitted}(x_n, \mathbf{play}),$$

*which says that if $x_0$ is Alice's boss, $x_1$ is $x_0$'s boss, ..., and $x_n$ is $x_{n-1}$'s boss, then $x_n$ may play.* ∎

While many policy bases that arise in practice have no more than one bipolar literal in each clause, we have found two relatively common situations in which this is not the case. The first is when policies refer to properties that are, intuitively, defined in the environment. The second is when the policy set includes both permitting and denying policies (that is, the set has policies with $\mathbf{Permitted}$ in the conclusion and policies with $\neg\mathbf{Permitted}$ in the conclusion).

To see why the bipolar restriction might be violated in the presence of definitions, consider a video store that has three types of customers: regular, gold, and platinum. Every adult member is permitted to send queries to the store's helpdesk, where adulthood is defined by the state in which the individual resides. In New York, an individual is an adult if she is over twenty-one years old. In Alaska, an

---

substitutes $\mathbf{f}(x)$ for $y$, instead of substituting **Alice** for $x$ and $\mathbf{f}(\mathbf{Alice})$ for $y$. So, the resolvent of $c$ and $c'$ is $\forall x (\mathbf{S}(\mathbf{f}(x)) \vee \mathbf{S}(\mathbf{g}(x)))$. (See [Nerode and Shore 1997] for details.)

individual is an adult if she is over eighteen. Formally,

$$p_1 = \forall x(\mathbf{Adult}(x) \wedge \mathbf{Member}(x) \Rightarrow \mathbf{Permitted}(x, \mathbf{query\ helpdesk}))$$
$$e_1 = \forall x(\mathbf{Over21}(x) \wedge \mathbf{InNY}(x) \Rightarrow \mathbf{Adult}(x))$$
$$e_2 = \forall x(\mathbf{Over18}(x) \wedge \mathbf{InAK}(x) \Rightarrow \mathbf{Adult}(x))$$
$$e_3 = \forall x(\mathbf{RegMember}(x) \Rightarrow \mathbf{Member}(x))$$
$$e_4 = \forall x(\mathbf{GoldMember}(x) \Rightarrow \mathbf{Member}(x))$$
$$e_5 = \forall x(\mathbf{PlatinumMember}(x) \Rightarrow \mathbf{Member}(x))$$

Roughly speaking, $e_1$ and $e_2$ define the notion of being an adult, while $e_3$, $e_4$, and $e_5$ define the notion of being a member. These definitions are used in $p_1$ to regulate who may send queries to the helpdesk. It is easy to see that $p_1$ has two bipolar literals in $p_1 \wedge e_1 \wedge \ldots \wedge e_5$, namely $\mathbf{Adult}(x)$ and $\mathbf{Member}(x)$ Therefore, the bipolar restriction does not hold in this example. More generally, if a policy $p$ mentions $k$ terms that are defined in the environment, then $p$ will include $k$ bipolar literals.

Definitions in this spirit arise frequently in government legislation, including the Social Security database and the Privacy Rule. Thus, handling definitions is a matter of practical importance. Perhaps the simplest approach is to rewrite the policy $p_1$ so as to replace the defined predicates in the antecedent by their definitions. This will result in an equivalent policy base with no bipolars. The effect of replacing $\mathbf{Adult}$ and $\mathbf{Member}$ by their definitions in our example is to replace $p_1$ by the six policies in $P_{NY} \cup P_{AK}$, where

$$P_{NY} = \{\forall x(\mathbf{Over21}(x) \wedge \mathbf{InNY}(x) \wedge \mathbf{Pr}(x) \Rightarrow \mathbf{Permitted}(x, \mathbf{query\ helpdesk})) :$$
$$\mathbf{Pr} \in \{\mathbf{RegMember}, \mathbf{GoldMember}, \mathbf{PlatinumMember}\}\}$$

$$P_{AK} = \{\forall x(\mathbf{Over18}(x) \wedge \mathbf{InAK}(x) \wedge \mathbf{Pr}(x) \Rightarrow \mathbf{Permitted}(x, \mathbf{query\ helpdesk})) :$$
$$\mathbf{Pr} \in \{\mathbf{RegMember}, \mathbf{GoldMember}, \mathbf{PlatinumMember}\}\}$$

Notice that there are no bipolars in $\bigwedge_{p \in P_{NY} \cup P_{AK}} p$ and the policies permit the same actions as $p_1 \wedge e_1 \wedge \ldots \wedge e_5$.

Our translation illustrates a potential problem with this approach: it can blow up the size of the policy set. Suppose that a policy $p$ has $m$ bipolar literals and that literal $i$ is defined using $c_i$ clauses. Rewriting would result in replacing policy $p$ by $c_1 \times \cdots \times c_m$ policies. Each of the new policies can also be longer than $p$, although the total length of each one can be no more than $|E_1|$, where $E_1$ is the first-order part of the environment. Is this so bad? Examples in the social security database and in the Privacy Rule suggest that typically $m$ is less than three and $i$ is less than five, in which case definitions do not significantly reduce the efficiency of our procedures.

In practice, we can often improve efficiency by removing definitions that are irrelevant when answering queries in a given environment. Continuing our earlier example, suppose that $E_0$ is the environment that results from Alice by presenting certificates that show she is a regular member who is over eighteen and in Alaska (i.e., $E_0 = \mathbf{RegMember}(\mathbf{Alice}) \wedge \mathbf{Over18}(\mathbf{Alice}) \wedge \mathbf{InNY}(\mathbf{Alice})$). It is easy to see that we can remove $e_1$, $e_4$, and $e_5$ without changing the set of permissions that are implied by the policy base. In practice, we believe that this single optimization will usually result in each $c_i$ being one (i.e., every predicate is defined by at most one clause), in which case our approach to handling definitions does not increase the number of policies mentioned in the query. As an aside, this optimization is one of many that are well-known in the theorem-proving community. We suspect that, by applying the appropriate optimizations, we can answer queries substantially faster than is indicated by the worst-case complexity results given in Theorem 4.7.

We next show how we can deal with policy bases that have both permitting and denying policies. This task would be easy if we could consider only the permitting policies (ignoring the denying policies) when determining if an action is permitted. Unfortunately, if we do this, then we might not answer queries correctly.

To see why, consider an environment $E$ that says "Alice is a student" and a policy set $\mathcal{P} = \{p_1, p_2, p_3\}$, where $p_1$ says "faculty members may chair committees", $p_2$ says "students may not chair committees", and $p_3$ says "anyone who is not a faculty member may take naps". We can write these policies as

$$p_1 = \forall x(\textbf{Faculty}(x) \Rightarrow \textbf{Permitted}(x, \text{chair committees})),$$
$$p_2 = \forall x(\textbf{Student}(x) \Rightarrow \neg\textbf{Permitted}(x, \text{chair committees})),$$
$$p_3 = \forall x(\neg\textbf{Faculty}(x) \Rightarrow \textbf{Permitted}(x, \text{nap})).$$

Clearly, $p_1$ and $p_3$ are permitting policies and $p_2$ is a denying policy. Because $p_1$ is equivalent to $\forall x(\neg\textbf{Permitted}(x, \text{chair committees}) \Rightarrow \neg\textbf{Faculty}(x))$, $p_1$ and $p_2$ together imply that no student is a faculty member. (Intuitively, students cannot be faculty members, because no one can be both permitted and not permitted to chair committees.) Because students are not faculty members, Alice, being a student, is not a faculty member and, by $p_3$, may take a nap. We cannot determine that Alice may nap if we consider only the permitting policies, because to derive the permission we need the environment fact that is implied by $p_1 \wedge p_2$.

If each fact implied by a permitting and denying policy together were derivable from either the environment or a single policy, then we could separate the permitting policies from the denying policies. Intuitively, this is because the interaction would not provide any information that was not already known. To formalize this intuition, note that each implied fact corresponds to a resolvent of a permitting and denying policy. In the previous example, the implied fact that students are not faculty members corresponds to the resolvent of $p_1$ and $p_2$, namely $\forall x(\textbf{Faculty}(x) \Rightarrow \neg\textbf{Student}(x))$. Therefore, if every resolvent of a permitting and denying policy is already implied by the environment or a single policy, then we can separate the policies. Continuing our example, we could separate the policies if the environment said that students were not faculty members. A closer analysis shows that we need to consider only those resolvents that are created by resolving on a literal that mentions **Permitted**.

To formalize all this, we need to discuss permitting and denying policies in a bit more detail. Note that a policy such as $\forall x(\textbf{Permitted}(\text{Alice}, a) \Rightarrow \textbf{Permitted}(\text{Bob}, a))$ is logically equivalent to both a permitting policy and a denying policy. (The denying policy is $\forall x(\neg\textbf{Permitted}(\text{Bob}, a) \Rightarrow \neg\textbf{Permitted}(\text{Alice}, a))$.) We say that a policy is *pure* if it is not logically equivalent to both a permitting and a denying policy. For example, policies that do not mention **Permitted** in the antecedent (which is the case for almost all the policies we have collected) are guaranteed to be pure.

THEOREM 4.5. *Suppose that $E$ is a standard environment, $P$ is a conjunction of pure permitting policies, and $D$ is a conjunction of (not necessarily pure) denying policies such that, for every resolvent $f$ created by resolving a conjunct of $P$ and a conjunct of $D$ on a literal that mentions* **Permitted**, *either $E \Rightarrow f$ is valid or $q \Rightarrow f$ is valid for some conjunct $q$ of $P \wedge D$. Then, for all terms $t$ and $t'$ of the appropriate sort, $E \wedge P \wedge D \Rightarrow$ **Permitted**$(t, t')$ is valid iff $E \wedge P \Rightarrow$ **Permitted**$(t, t')$ is valid.*

We can always add clauses to a policy base to obtain an equivalent policy base that satisfies the antecedent of Theorem 4.5. Therefore, the key question is not "how likely are these conditions to hold in practice", but "how many clauses are we going to have to add in practice so that these conditions hold". Example 4.4 shows that we may need to add an infinite number of policies to the set. However, in practice, policies are often simple. (Recall that a policy $p$ is simple if the antecedent of $p$ does not mention **Permitted**.) If every policy in a policy base is simple, then every resolvent is an environment fact and there is, at most, one resolvent per pair of permitting and denying policies. So, if the policy base mentions $n$ policies, all simple, then we can satisfy the antecedent of Theorem 4.5 by adding at most $n^2$ clauses to the environment.

Adding clauses to the environment, however, can have an unfortunate consequence. Suppose that $E$, $P$, and $D$ are as defined in Theorem 4.5 and $E'$ is $E$ extended so that the antecedent of Theorem 4.5 holds. Then the policy bases $E' \wedge P$ and $E' \wedge D$ might violate the bipolar restriction, even if $E \wedge P$ and $E \wedge D$ do not. To illustrate the problem, recall our earlier example in which policy $p_1$ says "faculty members may chair committees", $p_2$ says "students may not chair committees", and $p_3$ says "anyone who is not a faculty member may take a nap". Consider a policy set that consists of $p_1$, $p_2$, $p_3$, and a policy $p_4$ that says "anyone who is not a student may not enter the student-only website" $(\forall x(\neg \textbf{Student}(x) \Rightarrow \neg \textbf{Permitted}(x, \textbf{enter student site})))$. To satisfy the conditions of Theorem 4.5, we could add a clause $e$ to the environment that says "students are not faculty" $(e = \forall x(\textbf{Students}(x) \Rightarrow \neg \textbf{Faculty}(x))$. By Theorem 4.5, we can now separate the permitting and denying policies. However, determining if a permission is denied might be an intractable problem because the denying policies together with $e$ violate the bipolar restriction; $e$ has two bipolar literals in $e \wedge p_2 \wedge p_4$.

In this example, we can avoid the problem by satisfying the antecedent of Theorem 4.5 in another way. Rather than adding $e$ to the environment, we could replace $p_1$ by the policy $p_1' = \forall x(\textbf{Faculty}(x) \wedge \neg \textbf{Student}(x) \Rightarrow \textbf{Permitted}(x, \textbf{chair committees}))$, which says that faculty members who are not students may chair committees. Note that every clause in $p_1' \wedge p_3$ has at most one literal that is bipolar in $p_1' \wedge p_3$, and every clause in $p_2 \wedge p_4$ has at most one literal that is bipolar in $p_2 \wedge p_4$. So the antecedent of Theorem 4.5 holds and the resulting policy bases satisfy the bipolar restriction. We suspect that, in practice, a policy base $E \wedge P \wedge D$ either satisfies the bipolar restriction or can be converted to an equivalent policy base $E' \wedge P \wedge D$ that satisfies the antecedent of Theorem 4.5 and has the property that both $E' \wedge P$ and $E' \wedge D$ satisfy the bipolar restriction. We have not, however, done an extensive check.

Instead of adding these clauses to the environment automatically, it might be better to verify the changes with the policy maker. To see why, recall the two policies "faculty members may chair committees" and "students may not chair committees". We could satisfy the antecedent of Theorem 4.5 by adding the fact "no student is a faculty member" to the environment. But suppose that there is (or could one day be) a student who is also a faculty member. Then the policy maker may want to revise the policies to take this into account, rather than allowing the environment to (possibly) become inconsistent. In general, we expect that the additional facts needed to satisfy the antecedent of Theorem 4.5 will be ones that either the user would agree should have been there all along or are ones that should not be there and in fact suggest that the policies should be rewritten.

4.2.3  *Dealing With the Equality Restriction.* To explain how we can relax the equality restriction, we need two definitions. We say that a standard query $q$ of the form $E_0 \land E_1 \land P \Rightarrow \mathbf{Permitted}(t, t')$ is *equation-free* if no conjunct of $E_0 \land E_1 \land P$, when written in CNF, has a disjunct of the form $t = t'$. (Note that an equation-free query may mention equality in its antecedent, but only in the scope of negation when the antecedent is written in CNF. Thus, for example, the query $a \neq b \land (c = d \Rightarrow \mathbf{Permitted}(t, t')) \Rightarrow \mathbf{Permitted}(t, t')$ is equation-free.) It is easy to see that Theorem 4.1 applies to equation-free queries; it is only positive occurrences of $=$ that cause problems.

We can actually go slightly beyond equation-free queries. If $F_0$ is the conjunction of equality statements in $E_0$, then $q$ is *equality-safe* provided that $E_1 \land P$ (when written in CNF) has no clause with a disjunct of the form $t = t'$ and it is *not* the case that $F_0 \Rightarrow t = t'$ is valid, where $t$ and $t'$ are closed terms that appear in $E_0$ and either $t$ is a subterm of $t'$ or both $t$ and $t'$ mention function symbols. For example, $q$ is not equality-safe if $E_0$ includes the conjunct $\mathbf{c} = \mathbf{f}(\mathbf{c})$, the conjunct $\mathbf{f}(\mathbf{c}) = \mathbf{f}'(\mathbf{c}')$, or both $\mathbf{f}(\mathbf{c}) = \mathbf{c}'$ and $\mathbf{c}' = \mathbf{f}'(\mathbf{c}')$ (since these together imply $\mathbf{f}(\mathbf{c}) = \mathbf{f}'(\mathbf{c}')$).

Note that the notion of equality-safe is a generalization of equation-free; an equality-safe query can have some conjuncts in $E_0$ where equality does not appear in the scope of a negation, as long as not too much can be inferred from those equality statements. The following proposition shows that we can efficiently convert an equality-safe query $q$ to an equation-free query $q'$ such that $q$ is valid iff $q'$ is valid. Thus, we can determine the validity of an equality-safe query by first transforming it to an equation-free query and then applying the techniques discussed previously.

PROPOSITION 4.6. *If $q$ is an equality-safe standard query, then there is a standard query $q'$ of the form $E_0' \land E_1' \land P' \Rightarrow \mathbf{Permitted}(t, t')$ such that (a) $q$ is valid iff $q'$ is valid, (b) $q'$ is equation-free, and (c) $|q'| = O(|q||L_q'|)$, where $L_q'$ is the length of the longest term in $q$. Moreover, we can find such a $q'$ in time $O(|q|)$.*

Example B.8 in the appendix illustrates the procedure for converting $q$ to $q'$, and shows the problems that arise if we allow queries that are not equality-safe. The example also shows that the transformation procedure can increase the number of bipolar literals. Since we need to restrict the number of bipolars for tractability, our theorems must refer to the number of bipolars after the transformation. We say that $\ell$ and $\ell'$ are *unifiable relative to a set $E$ of equality statements* if there are variable substitutions $\sigma$ and $\sigma'$ such that it follows from $E$ that $\ell\sigma = \ell'\sigma'$. For example, $\mathbf{P}(\mathbf{a})$ and $\mathbf{P}(\mathbf{b})$ are unifiable relative to $\mathbf{a} = \mathbf{b}$, and $\mathbf{Permitted}(\mathbf{Alice}, \mathbf{nap})$ and **Permitted**(**wifeOf**($x$), **nap**) are unifiable relative to **Alice** = **wifeOf**(**Bob**). Similarly, we can talk about a literal $\ell$ being bipolar in a formula $f$ relative to $E$. If every conjunct in $E_1 \wedge P$ has at most one literal that is bipolar in $E_1 \wedge P$ relative to the equality statements in $E_0$, then after the transformation each conjunct will have at most one bipolar literal, which is what we need for tractability.

As we show in Theorem 4.7, we can handle equality-safe formulas. This suffices to handle the use of equality in all of the library and government policies that we collected, as well as the uses of equality in XrML and ODRL.

4.2.4   *The Effect of Unconstrained Variables.*   Let $q$ be a standard query of the form $E_0 \wedge E_1 \wedge P \Rightarrow$ **Permitted**($t, t'$) that satisfies the four restrictions of Theorem 4.1 (possibly relaxed as discussed in Sections 4.2.2 and 4.2.3). These restrictions essentially guarantee that (a) $q$ is valid if and only if there is a clause $c$ in $R(E_1 \wedge P)$ such that $E_0 \wedge c \Rightarrow$ **Permitted**($t, t'$) is valid and (b) $R(E_1 \wedge P)$ is relatively small. (This is made precise in Section 4.2.5.) The role of the variable restriction is to ensure that, for each $c$ in $R(E_1 \wedge P)$, we can quickly determine whether $E_0 \wedge c \Rightarrow$ **Permitted**($t, t'$) is valid. We now relax the variable restriction in a way that preserves this property.

Let $c$ be a conjunct of $E_1 \wedge P$. A variable $v$ is *constrained in $c$ relative to $q$* if $v$ appears as an argument to a literal that mentions **Permitted**, is a disjunct of $c$, and is not bipolar in $E_1 \wedge P$ relative to the equality statements in $E_0$. For example, consider the query "may Alice read file A" given that Alice is Ms. Jones, Alice may copy any file to any destination, and if Ms. Jones may copy a file to a destination, then she may read that file. We can write this query as

$$q = (\textbf{Alice} = \textbf{Ms. Jones}) \wedge p_1 \wedge p_2 \Rightarrow \textbf{Permitted}(\textbf{Alice}, \textbf{Read}(\textbf{file A})),$$

where

$p_1 = \forall x_1 \forall x_2 (\textbf{Permitted}(\textbf{Alice}, \textbf{copySrcDst}(x_1, x_2)))$, and
$p_2 = \forall x_1 \forall x_2 (\textbf{Permitted}(\textbf{Ms. Jones}, \textbf{copySrcDst}(x_1, x_2)) \Rightarrow \textbf{Permitted}(\textbf{Ms. Jones}, \textbf{Read}(x_1)))$.

Note that **Permitted**(**Ms. Jones**, **Read**($x_1$)) is the only literal in $p_1 \wedge p_2$ that is not bipolar in $p_1 \wedge p_2$ relative to **Alice** = **Ms. Jones**. It follows that no variable is constrained in $p_1$ relative to $q$; $x_1$ is constrained in $p_2$ relative to $q$; and $x_2$ is not constrained in $p_2$ relative to $q$.

If every literal in every conjunct $c$ of $E_1 \wedge P$ mentions at most one variable that is not constrained in $c$ relative to $q$, then it is not hard to show that every literal in every clause $c'$ in $R(E_1 \wedge P)$ mentions at most one variable that is not constrained in $c'$. (Recall that a variable is constrained in a clause $c$, as opposed to being constrained in $c$ relative to a query, if it appears in $c$ as an argument to **Permitted**). It turns out that this property is sufficient to ensure that we can quickly determine the validity of $E_0 \wedge c \Rightarrow$ **Permitted**($t, t'$) for each $c$ in $R(E_1 \wedge P)$. It can also be shown that if every conjunct $c$ of $E_1 \wedge P$ mentions at most $k$ variables that are not constrained in $c$ relative to $q$, then every clause $c$ in $R(E_1 \wedge P)$ mentions at most $2k$ variables that are not constrained in $c$. In this case, it can be shown that the validity of $E_0 \wedge c \Rightarrow$ **Permitted**($t, t'$) for a clause $c$ in $R(E_1 \wedge P)$ can be determined in time exponential in $2k$. (All these claims are made precise in Theorem 4.7.) It follows that if $k$ is less than three, which is likely to be the case in practice, then we can remove the variable restriction entirely and still answer queries in a reasonable period of time.

### 4.2.5 *Putting It All Together.*
With all this machinery, we can finally define the fragment of first-order logic that we believe to be appropriate for expressing policies. *Lithium* consists of all equality-safe standard queries $E_0 \wedge E_1 \wedge P \Rightarrow \textbf{Permitted}(t, t')$ such that every conjunct in $E_1 \wedge P$ has at most one literal that is bipolar in $E_1 \wedge P$ relative to the equality statements in $E_0$.

We can quickly determine whether a query $q$ is in Lithium. To determine if $q$ is equality-safe, we create equivalence classes for the terms in $E_0$, which takes linear time, and then verify that each class has at most one term that is not a constant, which also takes linear time. To determine if the bipolar restriction holds, we choose a term from each equivalence class to represent the class (choosing a term that is not a constant if possible) and replace each term in the query by its representative. The bipolar restriction holds if each conjunct $c$ of $E_1 \wedge P$ has at most one literal that is bipolar in $E_1 \wedge P$, which we can check in quadratic time.

The discussion in Sections 4.2.1, 4.2.2, and 4.2.3 suggests that queries in Lithium are tractable. The following theorem makes this precise. We prove the theorem only for equation-free Lithium queries, but by Proposition 4.6, it applies to equality-safe queries as well (although the complexity statements would have to be changed to take into account the possible increase in size when converting from equality-safe to equation-free queries).

Let $L_f$ be the length of the longest clause in a CNF formula $f$, and let $L'_f$ be the length of the longest term in $f$.

THEOREM 4.7. *The validity of an equation-free Lithium query $q = E_0 \wedge E_1 \wedge P \Rightarrow$* $\textbf{Permitted}(t, t')$ *with $m$ terms in $E_0$ can be determined in time $O((|E_0| + T|E_1 \wedge P|^2) \log |E_0|)$, where $T = m L_{E_1 \wedge P} L'_{E_1 \wedge P} |\textbf{Permitted}(t, t')|$ if every literal in every conjunct $c$ of $E_1 \wedge P$ mentions at most one variable that is not constrained in $c$ relative to $q$; otherwise, $T = m^{2k} L_{E_1 \wedge P} L'_{E_1 \wedge P} |\textbf{Permitted}(t, t')|$, where every conjunct $c$ of $E_1 \wedge P$ has at most $k$ variables that are not constrained in $c$ relative to $q$.*

Theorem 4.7 shows that Lithium is tractable. Is it sufficiently expressive? The bipolar restriction holds in all of the applications that we considered, provided that definitions and mixed policy sets are handled as described in Section 4.2.2. We believe that our examples are representative, and that in fact the restriction will hold in practice. The restriction to equality-safe queries is likely to hold for applications that do not include a threshold policy; that is, a policy of the form "if $k$ instances of $P$ hold then subject $t_1$ may do action $t'$". If an application includes threshold policies, then the restriction is still likely to hold provided that the environment stores the number of relevant instances of $P$ that hold rather than the instances themselves. For example, the threshold policy "if two people blackball Alice, then she may not join the club" can be written in Lithium if the environment stores the number of people who blackball Alice (e.g., $\textbf{numOfBlackballers} = 2$), instead of who blackballs Alice (e.g., $\textbf{Blackballs}(\textbf{Bob}, \textbf{Alice}) \wedge \textbf{Blackballs}(\textbf{Carol}, \textbf{Alice})$).

## 5.    CONSISTENCY

Recall that a policy set is inconsistent if it both permits and forbids the same action. By detecting inconsistencies, we can warn policy writers that their policies probably do not match their intentions. We expect that this ability will be particularly important if the policy set is large or if it is created and maintained by more than one person. In addition, we can verify that a policy base $P$ is consistent with a policy base $P'$ by checking that $P \cup P'$ is consistent. For example, suppose that access to a patient's medical file is regulated by the hospital's policies and state law. If the union of the two policy bases is consistent, then the hospital's policies do not contradict state law. (Note that the converse is not necessarily true.)

Clearly, $E \wedge P$ is not consistent iff both $E \wedge P \Rightarrow \mathbf{Permitted}(c, c')$ and $E \wedge P \Rightarrow \neg\mathbf{Permitted}(c, c')$ are valid, for arbitrary constants $c$ and $c'$. Thus, if the two queries are in Lithium, then we can apply our previous techniques to show that we can efficiently check consistency. However, we can say even more. If the condition of Theorem 4.5 (or the corresponding condition for determining prohibitions) is met, then we automatically have consistency, provided that $E$ is consistent.

THEOREM 5.1. *Suppose that $E$ is an environment, $P$ is a conjunction of pure permitting policies, and $D$ is a conjunction of (not necessarily pure) denying policies such that the antecedent of Theorem 4.5 holds. Then $E \wedge P \wedge D$ is satisfiable iff $E$ is satisfiable.*

Thus, in addition to making it feasible to check the consequences of policies, our conditions essentially prevent users from writing inconsistent policies. This is a major benefit of adhering to these restrictions!

## 6.    USABILITY

In this section, we consider ways to make Lithium accessible to people who are not conversant with first-order logic. The restrictions on bipolars and equality in Lithium might be difficult to explain to non-logicians, but we suspect that teaching people to write standard queries can be done quickly, particularly if syntactic sugar is used to help the medicine go down.

We are currently designing usability tests to verify that computer programmers can learn to translate English sentences to standard (sugared) queries quickly. The "sugaring" involves, for example, rewriting "$\forall x_1 \ldots \forall x_n (\ell_1 \wedge \ldots \wedge \ell_k \Rightarrow \ell_{k+1})$" as "$\mathbf{type_1}\ x_1; \ldots, \mathbf{type_1}\ x_n; \mathbf{if}\ \ell_1\ \mathbf{and}\ \ldots\ \mathbf{and}\ \ell_k\ \mathbf{then}\ \ell_{k+1}$", where $\mathbf{type_i}$ is the sort of variable $x_i$. We are focusing on programmers because, if this community can read and write queries, then they can build user interfaces for other communities, along with translators that convert user input to queries. Of course, input entered through a user interface can also be translated directly to a (non-sugared) standard query. For example, it should be possible to write a form-based interface that allows users to enter queries, which can then be translated directly to Lithium. Such a form-based interface was sketched in the conference version of the paper [Halpern and Weissman 2003]. We have not pursued it because we feel it is better to write an interface for programmers.

The key question is how we should explain the bipolar and equality restrictions to policy writers. One option is to define a fragment of Lithium that is easy to explain to non-logicians and fairly expressive. For example, let $S$ be the set of standard queries in which the environment is basic (a conjunction of **Permitted**-free literals), the policies are simple (**Permitted** is not mentioned in the antecedent), and the antecedents of policies are negation-free. It is easy to see that every query in $S$ is in Lithium. Another example is Rosetta [Weissman and Lagoze 2004]. Rosetta is a fragment of (somewhat stilted) English in which queries can be written. All queries that can be expressed in Rosetta are guaranteed to be convertible to Lithium. Finally, graphical interfaces can be designed in such a way that every query written using the interface can be translated to Lithium. We conjecture that the Information Rights Management system that is part of Microsoft's Office, Professional Edition 2003 [Microsoft 2003] is an example of this approach, although we have not verified that all policies written through these interfaces satisfy the bipolar and equality restrictions. In short, we believe that, for many applications, there is a fragment of Lithium that is both sufficiently expressive and accessible to users with minimal training. Which fragment is appropriate depends on the capabilities of the users and the needs of the application.

Another approach is to give policy writers guidelines and tools to help them write policy bases that satisfy our requirements. For example, we might suggest that policy writers try to minimize their use of negation, equality, and universal formulas in the environment. We can provide tools to check if proposed policy bases are likely to lead to queries that are in Lithium. In practice, we expect policy writers to define the universal formulas in the environment and the policies (i.e., $E_1 \wedge P$); individuals then present certain credentials (i.e., $E_0$) along with a request (i.e., **Permitted**$(t, t')$). In this setting, we can check if the bipolar restriction and equality restriction are satisfied by $E_1 \wedge P$ and, if so, we can conclude that every query of the form $E_0 \wedge E_1 \wedge P \Rightarrow$ **Permitted**$(t, t')$ is in Lithium provided that $E_0$ is equality-free. This allows us to identify potential problems at "compile time" and alert the policy writer, who might then choose to change the policies and environment to more closely adhere to the guidelines.

Perhaps the simplest solution is to not do anything at all. We believe, and have argued throughout this paper, that queries in practice are likely to be in Lithium. So users might not need to understand the restrictions on bipolar literals and equality, because they will naturally write queries that satisfy our requirements. We can build a verifier to check that a user's query is either in Lithium or can be converted to Lithium using the techniques discussed in Section 4.2.2. If a query is in Lithium, then the user is assured that her question will be answered efficiently. Otherwise, the verifier issues a warning. The warning could be ignored since our algorithm for answering queries might still run efficiently or, since warnings are likely to be rare, an expert could be consulted.

We expect that all of these strategies will allow naive users to express their queries in Lithium easily. It is up to the application developers to decide which approach is best in their setting.

## 7. RELATED WORK

There has been a great deal of work on policy languages. Since we cannot hope to review all of the work in only a few pages, we restrict our attention to some of the best-known approaches and to those that seem most similar to Lithium.

The classic approach in the Computer Science community is arguably the one taken by UNIX. Every policy in UNIX can be expressed as a formula of the form $\forall x (\mathbf{R}(x, r) \Rightarrow \mathbf{Permitted}(x, \mathbf{act}(r)))$, where $\mathbf{R} \in \{\mathbf{User}, \mathbf{Group}, \mathbf{Other}\}$, $\mathbf{act} \in \{\mathbf{read}, \mathbf{write}, \mathbf{execute}\}$, and $r$ is a constant typically representing a file or directory. The corresponding environment can be written as a conjunction of ground literals. It is easy to see that every query in UNIX can be written in Lithium. However, the UNIX approach to answering a query is somewhat different than that taken by Lithium. UNIX assumes that every action not explictly permitted is forbidden. Thus, with an empty environment ($true$), the UNIX response to the query $\neg\mathbf{Permitted}(\mathbf{Alice}, \mathbf{read}(\mathbf{file\ f}))$ would be yes, while the Lithium response would be no (it does not logically follow from $true$ that Alice is not permitted to edit file f. We can modify Lithium to give the same answers to queries as UNIX, simply by saying that the answer to a query of the form $\neg\mathbf{Permitted}(t, t')$ given a policy base $b$ is yes iff $b \Rightarrow \mathbf{Permitted}(t, t')$ is not acceptably valid; that is, a prohibition holds if and only if the corresponding permission does not. Since we know how to determine whether a permission holds, we can determine if a prohibition holds according to the revised definition. This modified version of Lithium can also capture the way policies are evaluated using access control lists. We believe that we can also capture whether a permission follows in SPKI/SDSI [Ellison et al. 1999a; 1999b] from a collection of certificates in Lithium, although we have not checked the details.

Perhaps the most talked-about policy language in industry today is the XML-based language XACML [Moses 2005]. Every XACML query can be written as a standard query in which all policies are simple (the antecedents of policies are $\mathbf{Permitted}$-free) and the environment is basic (a conjunction of $\mathbf{Permitted}$-free ground literals). There are two significant differences between XACML and Lithium. The first is that users of XACML are expected to provide an algorithm for determining whether a permission is granted, denied, or unregulated by a policy base, as a function of whether the permission is granted, denied, or unregulated by the individual polices in that policy base. For example, the *deny-overrides* algorithm (which is one of the built-in algorithms provided by XACML) says a permission is denied if it is denied by any single policy, is permitted if it is not denied by any single policy and is permitted by at least one, and is unregulated otherwise. Lithium essentially allows only one algorithm, which is logical consequence (a choice which cannot in fact be expressed in XACML, since it may depend on the interaction between the policies in a policy base). We could, of course, modify the way Lithium handles queries to match any particular algorithm, although doing this may result in losing many of the unique features of Lithium.

The second key difference between XACML and Lithium is the treatment of negation. In XACML, the semantics of negation is somewhat nonstandard. For example, in XACML, the policies "if Alice is good, then she may play" and "if Alice is not good, then she may play" together do not necessarily imply that Alice may play. The policies imply the permission only if the environment says either that Alice is good or that she is not good. So, given a set of XACML policies, we can replace every literal of the form $\neg\mathbf{R}(t_1, \ldots, t_n)$ by $\mathbf{NotR}(t_1, \ldots, t_n)$, where $\mathbf{NotR}$ is a fresh predicate symbol, without changing the meaning of the policies. Thus, although XACML seems to allow the unrestricted use of negation, it is actually less expressive than Lithium in its use of negation. Moreover, we believe that the nonstandard usage of negation may well confuse users.

Another XML-based language that has received widespread support in industry is XrML [ContentGuard 2001]. XrML and Lithium are incomparable in expressive power. XrML is less expressive in that it does not allow negation. This means in particular that it cannot express denying policies and cannot capture a policy that grants a permission based on whether a condition does *not* hold. In addition, the conclusion of every environment fact that is not a ground literal is of the form $R(p)$, where $R$ is a unary predicate symbol and $p$ is a principal. On the other hand, XrML is more expressive than Lithium in that a policy can grant a permission based on the answers to various queries. For example, in XrML, Alice's babysitter can write the policy "Alice is permitted to do some action $a$ if the permission follows from her mother's policies and from her father's policies". We can extend Lithium to include such policies as well. Let Lithium$^+$ be Lithium extended with a $\mathsf{Val}$ operator, where $\mathsf{Val}(\varphi)$ is true if $\varphi$ is valid. We can write the babysitter's policy in Lithium$^+$ as $\forall x (\mathsf{Val}(E_M \wedge P_M \Rightarrow \mathbf{Permitted}(\mathbf{Alice}, x)) \wedge \mathsf{Val}(E_D \wedge P_D \Rightarrow \mathbf{Permitted}(\mathbf{Alice}, x)) \Rightarrow \mathbf{Permitted}(\mathbf{Alice}, x))$, where $E_M \wedge P_M$ and $E_D \wedge P_D$ are the policy bases of Alice's mother and father respectively. We can place restrictions on Lithium$^+$ similar in spirit to those on Lithium to ensure that it is tractable, yet expressive enough to capture the policies that users want in practice; see [Halpern and Weissman 2004] for details.

The policy languages that are perhaps closest in spirit to Lithium are the approaches that are based on some variant of Datalog. Examples of such languages include Delegation Logic [Li et al. 2003], the RT (Role-based Trust-management) framework [Li et al. 2002], Binder [DeTreville 2002], SD3 [Jim 2001], FAF (Flexible Authorization Framework) [Jajodia et al. 2001], and Cassandra [Becker and Sewell 2004]). Datalog is an efficient well-understood reasoning engine that is restricted to function-free negation-free Horn clauses; these restrictions are made to ensure tractability. The variants, such as *safe stratified Datalog* [Garcia-Molina et al. 2002] or *Datalog with constraints*, allow limited use of functions and negation while preserving tractability.

The main difference between Lithium and these Datalog-based languages is in the use of functions and negation. There are relatively few policy languages that include functions symbols, but those that do (e.g. [Bertino et al. 1998; Li and Mitchell 2003; Becker and Sewell 2004]) seem to favor Datalog with constraints. By using this variant of Datalog, many structured resources, such as directories, can be expressed using functions. However, function symbols may not appear in intentional predicates (predicates whose relations are computed by applying Datalog rules, as opposed to being stored in a database). For example, the policy "every authorized individual may copy a classified file from one secure server to another" when written as

$\forall x_1 \dots \forall x_4 (\mathbf{Auth}(x_1) \wedge \mathbf{Classified}(x_2) \wedge \mathbf{Secure\_Server}(x_3) \wedge \mathbf{Secure\_Server}(x_4)$
$\Rightarrow \mathbf{Permitted}(x_1, \mathbf{copySrcDst}(x_2, x_3, x_4)))$

is not in Datalog with constraints. Also, for tractability, additional restrictions are often made. For example, Li and Mitchell [2003] do not allow formulas in constraints to have more than one variable and Becker and Sewell [2004] require that every argument of a function in a query be variable-free.

There are a number of policy languages that allow a limited use of negation. Jajodia, Samarati, Sapino, and Subrahmanian [2001] base their policy language on Datalog with negation, which is a variant of Datalog that allows unrestricted use of negation in the body of rules. Datalog with negation is tractable because it makes the *closed-world assumption*: if we cannot prove that a positive literal is true, we take it to be false. Unfortunately, the closed-world assumption can lead to unintuitive (and probably unintended) results. For example, consider the policy "if Alice does not have bad credit, then she may apply for a loan", and suppose that the reasoning engine determines whether an individual has bad credit by reviewing her credit report. If Alice has bad credit and does not present her credit report, then a reasoning engine that makes the closed-world assumption will incorrectly assume that Alice does not have bad credit and thus will allow her to make a loan application.

Several policy languages (e.g. [DeTreville 2002; Li et al. 2003; Li et al. 2002; Jim 2001]) are based on safe stratified Datalog. Safe stratified Datalog allows some use of negation in the body of rules and does not make the closed world assumption. However, the restrictions on negation still prevent it from capturing some permitting policies of interest. For example, the policy

$$\forall x(\neg \textbf{BadCredit}(x) \Rightarrow \textbf{Permitted}(x, \textbf{apply for loan}))$$

(anyone without bad credit may apply for a loan) cannot be expressed. More importantly, denying policies cannot be expressed in safe stratified Datalog because the language does not allow negation in the conclusion of rules.

This limitation may not seem to be particularly troublesome. After all, the standard approach is to assume that every permission not explicitly granted is denied. (For example, this is done in relational databases [Griffiths and Wade 1976], almost all of the Datalog-based languages, UNIX, SPKI/SDSI [Rivest and Lampson 1996; Ellison et al. 1999a; 1999b], and KeyNote [Blaze et al. 1996].) However, in many contexts, it is difficult to believe that policymakers really want to forbid *every* action that they do not explicitly permit, so there is a mismatch between a policymaker's intentions and the interpretation of the policy base. This becomes a problem when different policymakers want to compare policy bases or combine them. The following examples illustrate the concern.

EXAMPLE 7.1. *Suppose that a hospital wants to verify that its policies comply with federal regulations; that is, the hospital wants to check that, if the government permits an action, then the hospital permits it and, if the government forbids an action, then the hospital forbids it. If the policies are written in a language that captures only permissions, assuming all other actions are forbidden, then compliancy checking is essentially impossible. In particular, if the hospital permits any action that is not regulated by the government (e.g., nurses may park in Lot A, all staff are welcome to drink the coffee in the lounge), then the hospital will appear to be non-compliant because it permits an action that is not explicitly permitted by the government and, thus, is implicitly forbidden. In short, because we cannot* *distinguish forbidden actions from unregulated ones, compliance checking reduces to determining whether one policy set is essentially identical to another.* ∎

EXAMPLE 7.2. *Consider a group of libraries that want to merge their policies so that patrons are governed by the same regulations, regardless of which library they visit. When merging the policy sets, we clearly want to detect conflicts (e.g. one library lets minors check out adult books and another does not). Unfortunately, if a language can state only what is permitted, then this will be impossible. If we put the permitting policies from each library into one large set, then that set will be consistent (it is satisfied in the model that permits everything), regardless of which policies are in the set. Alternatively, we could require that no library permits an action that another forbids (which is what we want to do) under the assumption that every unregulated action is forbidden. It is not hard to see that this approach will always detect a conflict between sets of library policies unless the sets are essentially identical.* ∎

The issues involved with comparing and merging policy bases have by and large been ignored, but we believe they will become increasingly significant. It seems unlikely that a policy language will be able to support these features unless the language can express both permitting and denying policies.

Although we do not know of a Datalog variant that allows negation in the conclusions of rules (thereby allowing denying policies), some languages seem to capture something comparable. For example, in FAF, actions are either positive or negative; the statement "principal $p$ can do negative action $act$" means $p$ is forbidden to do $act$. Another option in the same spirit is to have the predicate symbol **Forbidden** in the language, in addition to **Permitted**. A consequence of this approach is that it is not logically inconsistent for an action to be both permitted and forbidden. (Note that this is also the case for XACML, due to its nonstandard interpretation of negation.) To handle inconsistencies, FAF expects the policy writer to create overriding policies such as "if an action is both permitted and forbidden, then it is forbidden". If an inconsistency is detected when answering a query, then the overriding policy is applied. Similar approaches are taken by Chomicki et al. [2000] and Ioannides and Selis [1992]. The main problem with capturing prohibitions in this way is that the answers to queries might not match a policy writer's expectations. Policy writers typically do not intend to write policies that both permit and forbid the same action. Rather than identifying such policies and alerting the policy writer, potential errors are patched with overriding policies. In addition, these overriding policies are required even for consistent policy bases, which seems rather burdensome.

Lithium deals with this issue in what is arguably a better way. Given a policy base written in Lithium, we can detect conflicts and determine why they occur at "compile time" rather than at "run time", when a particular query is evaluated. Using this information, the policy writer can modify the environment and policies to more closely match her intentions. Moreover, if the antecedent of Theorem 5.1 holds, then the policy base is inconsistent if and only if the environment alone is inconsistent. Thus, we can often determine when there is no conflict that needs to be addressed.

The use of function symbols and negation is not the only difference between Lithium and other policy languages. Unlike Lithium, many languages have explicit support for *groups* and *roles*. A group is a set of subjects such that if a group has a property, then every member of the group has the property (cf. [Abadi et al. 1993; Jajodia et al. 2001]). In role-based access control models [Ferraiolo et al. 1999; Hitchens and Varadharajan 2001; Li et al. 2002; Sandhu et al. 1996] roles are an intermediary between individuals and rights. More specifically, an individual obtains a right by assuming a role that is associated with that right. For example, Alice may need to assume the role of Department Chair in order to obtain the budget.

Predicate symbols can be used to capture groups and roles in first-order logic. For example, if we want to say that Alice is a member of the faculty and any faculty member may chair committees, then we can represent the group using the predicate **Faculty**. The environment fact is encoded as **Faculty**(**Alice**); the policy is then

$$\forall x(\textbf{Faculty}(x) \Rightarrow \textbf{Permitted}(x, \textbf{chair committees})).$$

Similarly, the policy "Alice, acting as the Department Chair, may sign the budget" can be written as

$$\textbf{Dept\_Chair}(\textbf{Alice}) \Rightarrow \textbf{Permitted}(\textbf{Alice}, \textbf{sign the budget}).$$

The fact **Dept\_Chair**(**Alice**) would be added to the environment when Alice assumes the role and would be removed when she relinquishes it. Alternatively, we could add a sort *Roles* to our logic along with the predicate **As** (as suggested by Lampson, Abadi, Burrows, and Wobber [1992]), where **As**$(e, r)$ means that entity $e$ is acting as role $r$ (in other words, $e$ has assumed role $r$). Continuing our example, "Alice, acting as the Department Chair, may sign the budget" could be written in the logic as **As**(**Alice**, **Dept\_Chair**) $\Rightarrow$ **Permitted**(**Alice**, **sign the budget**). The second encoding for roles may be more in keeping with the spirit of the role-based model, but we believe that both approaches are reasonable (and our results apply to both choices). In short, Lithium supports groups and roles implicitly.

Lithium, as well as the Datalog variants, all use a fragment of first-order logic to express policies. Other approaches use a modal logic. Formal work on deontic logic (the logic of "obligation" and "permission") goes back to von Wright [1951]. Glasgow, MacEwen, Panangaden [1992] were the first to base a formal logic of security on deontic logic. The logic of access control consider by Lampson et al. and Abadi et al. [Lampson et al. 1992; Abadi et al. 1993] can also be viewed as a modal logic, with a *says* operator. These approaches can be translated into first-order logic, but they have features that take them beyond Lithium. For example, Abadi et al. have a calculus of principals; Glasgow, McEwen, and Panangaden deal with obligation as well as permission. We believe that many of these features could be added to Lithium, but we have not explored this issue.

The KeyNote system [Blaze et al. 1998], which is based on PolicyMaker [Blaze et al. 1996], is more flexible than Lithium in that the application can invoke policies written in a number of different languages. These are programs that determine if a policy applies to a request and a requestor. Because KeyNote essentially views these programs as black boxes, it is quite limited in its ability to reason about policies. As discussed by Blaze, Feigenbaum, and Strauss [1998], the system needs to put restrictions on the programs to ensure correct analysis. This is in fact done in KeyNote, but at the price of a substantial reduction in the expressive power of the language.

Finally, we remark that the design of Lithium was heavily influenced by the work of Halpern, van der Meyden, and Schneider [1999]. They identify some key issues that must be addressed when developing a policy language, evaluate various solutions that have been proposed in the literature, and recommend directions for future research. Our design incorporates three of their suggestions. In particular, we write policies in first-order logic; define sorts for principals, actions, and time; and use a **Permitted** predicate that takes an individual and an action argument. (This usage of **Permitted** is much in the spirit of how it is used in modal deontic logic.)

## 8.  CONCLUSION

We have presented a fragment of first-order logic called Lithium that seems well-suited to reasoning about policies. Unlike previous approaches, Lithium allows nearly unrestricted use of function symbols while still preserving tractability. Moreover, Lithium can express prohibitions explicitly, making it possible to capture the merger of policies.

To be of practical use, a policy language must be accessible to non-experts with minimal training and it must be sufficiently expressive to capture the policies of real applications. To make Lithium more accessible, we have created a front end for it called Rosetta [Weissman and Lagoze 2004], and are currently designing appropriate usability tests. Whether a language is sufficiently expressive is obviously an empirical question: it depends on what people want to write. We have collected numerous policies and verified that they are all expressible in Lithium. Although this is certainly not a proof of anything, it does increase our confidence in the adequacy of Lithium's expressive power.

In future work, we would like to extend Lithium to reason about policies that change over time. We are also exploring whether a hybrid of Lithium and a Datalog-based fragment can allow a further increase of expressive power without sacrificing tractability.

## Appendix

## A.  PROOFS FOR SECTION 3

The following lemma is the key to proving Theorems 3.1 and 3.3.

LEMMA A.1. *Let $\mathcal{L}_0'$ be a set of closed formulas with no constant symbols whose only predicate symbol is* **Permitted**. *Let $\mathcal{L}_0''$ be the set of closed formulas of the form*

$$(f \Rightarrow \textbf{Permitted}(c, c')) \Rightarrow \textbf{Permitted}(c, c'),$$

*where $c$ and $c'$ are constants of the appropriate sorts and $f \in \mathcal{L}_0'$. If the validity problem for $\mathcal{L}_0'$ is undecidable, then the validity problem for $\mathcal{L}_0''$ is undecidable.*

PROOF. We reduce the validity problem for $\mathcal{L}_0'$ to the validity problem for $\mathcal{L}_0''$. Straightforward manipulations show that $(f \Rightarrow \textbf{Permitted}(c, c')) \Rightarrow \textbf{Permitted}(c, c')$ is equivalent to $f \vee \textbf{Permitted}(c, c')$. Clearly, if $f \vee \textbf{Permitted}(c, c')$ is not valid, then $f$ is not valid. Suppose that $f \vee \mathbf{Permitted}(c, c')$ is valid. Since $f$ does not mention a constant symbol, $c$ and $c'$ do not appear in $f$, so $f \vee \forall x \forall y \mathbf{Permitted}(x, y)$ is valid. It follows that $f$ is valid iff $f$ is true in all models $m$ that satisfy $\forall x \forall y \mathbf{Permitted}(x, y)$. To determine whether $f$ is true in $m$, let $f'$ be the result of replacing all occurrences of $\mathbf{Permitted}(x, y)$ by $true$. Clearly $f$ is true in $m$ iff $f'$ is true in $m$. Since $f'$ has no nonlogical symbols, $f'$ is true in $m$ iff $f'$ is valid. Moreover, the validity of $f'$ is easy to determine. □

THEOREM 3.1. *Let $\mathcal{L}_0$ be the set of closed formulas of the form*

$$(f \Rightarrow \mathbf{Permitted}(c, c')) \Rightarrow \mathbf{Permitted}(c, c'),$$

*where $c$ and $c'$ are constants of the appropriate sorts, $f$ has a single alternation of quantifiers, and the only nonlogical symbol in $f$ is $\mathbf{Permitted}$. The validity question for $\mathcal{L}_0$ is undecidable.*

PROOF. Let $\mathcal{L}_0^A$ be the set of closed formulas that have a single alternation of quantifiers and whose only nonlogical symbol is $\mathbf{Permitted}$. The proof follows from Lemma A.1, where we take $\mathcal{L}_0'$ to be $\mathcal{L}_0^A$, because the validity problem for $\mathcal{L}_0^A$ is undecidable [Börger et al. 1997]. □

THEOREM 3.3. *Let $\mathcal{L}_1$ be the set of closed formulas of the form*

$$\forall x_1 \forall x_2 (f \Rightarrow \mathbf{Permitted}(c, c')) \Rightarrow \mathbf{Permitted}(c, c'),$$

*where $c$ and $c'$ are constants of the appropriate sort and $f$ is a quantifier-free formula whose only nonlogical symbols are $\mathbf{Permitted}$ and a unary function. The validity problem for $\mathcal{L}_1$ is undecidable.*

PROOF. Let $\mathcal{L}_1^A$ be the set of closed formulas of the form $\exists x_1 \exists x_2 f$, where $f$ is a quantifier-free formula whose only nonlogical symbols are $\mathbf{Permitted}$ and a unary function. Because the validity problem for $\mathcal{L}_1^A$ is undecidable [Börger et al. 1997], it follows from Lemma A.1 that the validity problem for the set of formulas of the form

$$(\exists x_1 \exists x_2 f \Rightarrow \mathbf{Permitted}(c, c')) \Rightarrow \mathbf{Permitted}(c, c') \tag{1}$$

is undecidable. Standard manipulations show that a formula of the form (1) is equivalent to

$$\forall x_1 \forall x_2 (f \Rightarrow \mathbf{Permitted}(c, c')) \Rightarrow \mathbf{Permitted}(c, c').$$

It follows that the validity problem for $\mathcal{L}_1$ is undecidable. □

THEOREM 3.4. *Let $\Phi$ be a vocabulary that contains $\mathbf{Permitted}$, constants $c$ and $c'$ of sorts Subjects and Actions, respectively, and possibly other predicate and constant symbols (but no function symbols). Assume that there is a bound on the arity of the predicate symbols in $\Phi$ (that is, there exists some $N$ such that all predicate symbols in $\Phi$ have arity at most $N$). Finally, let $\mathcal{L}_2$ be the set of all closed formulas in $\mathcal{L}^{fo}(\Phi)$ of the form $E \wedge p_1 \wedge \ldots \wedge p_n \Rightarrow \mathbf{Permitted}(c, c')$ such that $E$ is a conjunction of quantifier-free and universal formulas and each policy $p_1, \ldots, p_n$ has the form $\forall x_1 \ldots \forall x_m (f \Rightarrow \mathbf{Permitted}(t_1, t_2))$, where $t_1$ and $t_2$ are terms of the appropriate sort and $f$ is quantifier-free.*

(a) *The validity problem for $\mathcal{L}_2$ is in $\Pi_2^P$.*

(b) *If $\mathcal{L}_3$ is the set of formulas in $\mathcal{L}_2$ in which every policy's antecedent is a conjunction of literals, then the validity problem for $\mathcal{L}_3$ is $\Pi_2^P$ hard.*

(c) *If $\mathcal{L}_4$ is the set of $\mathcal{L}_2$ formulas in which $E$ is quantifier-free, then the validity problem for $\mathcal{L}_4$ is both NP-hard and co-NP hard.*

PROOF. For part (a), straightforward manipulations show that each formula $h$ in $\mathcal{L}_2$ is equivalent to a closed formula of the form $g = \exists x_1 \ldots \exists x_k g'$, where $g'$ is a quantifier-free formula in $\mathcal{L}^{fo}(\Phi)$. Moreover, $|g|$ is polynomial in $|h|$. Suppose that $g$ mentions $n$ distinct constant symbols. Let $\mathcal{M}$ be the class of models whose domain size is at most $\max(n, 1)$. We claim that (1) $g$ is valid iff it is true in every model in $\mathcal{M}$, and (2) the problem of determining if $g$ is true in every model $m \in \mathcal{M}$ is in $\Pi_2^P$.

For part (1), the "only if" direction is trivial. To prove the "if" direction, suppose by way of contradiction that $g$ is true in every model in $\mathcal{M}$ and $g$ is not true in a model $m$ with domain $D$ and interpretation $I$. Let $D' = \{I(c) \mid c$ is a constant in $g\}$ if $g$ mentions at least one constant, and $D' = \{d\}$ for some fixed element $d \in D$ if $g$ does not mention any constants. Let $I'$ be the interpretation such that $I'(c) = I(c)$ if $I(c) \in D'$, $I'(c) = d'$ for some fixed $d \in D'$ if $I(c) \notin D'$, and $I'(R) = D'^k \cap I(R)$ for each $k$-ary predicate $R$ in $\Phi$. Let $m'$ be the model with domain $D'$ and interpretation $I'$. Notice that $m'$ is in $\mathcal{M}$. By assumption, $m'$ satisfies $g$, so there are domain elements $d_1, \ldots, d_k$ in $D'$ such that by interpreting $x_i$ as $d_i$ for $i = 1, \ldots, k$, $m'$ satisfies $g'$. Under the same interpretation of $x_1, \ldots, x_k$, $m$ satisfies $g'$. Therefore $m$ satisfies $g$, and we have the desired contradiction.

For part (2), first note that $g$ is true in all models in $\mathcal{M}$ iff it is true in all models with domain $\{1, \ldots, m\}$ for each $m \le max(n, 1)$, since every model in $\mathcal{M}$ is isomorphic to one with domain $\{1, \ldots, m\}$ for $m \le \max(n, 1)$. The truth of $g$ in such a model depends only on the the interpretation of the constant and predicate symbols that actually appear in $g$. Let a *restricted interpretation* be one that interprets only the symbols that appear in $g$. Because there are $m^k$ interpretations of a $k$-ary predicate, and the arity of predicates in $g$ is bounded in a domain of size $m$, the number of restricted interpretations is polynomial in $|g|$. It clearly can be determined in time polynomial in $|g|$ if the formula $g'$ is true under a given restricted interpretation in a model with domain $\{1, \ldots, m\}$. Thus, determining if $g$ is true in such a model is in NP (since it involves guessing an interpretation of $x_1, \ldots, x_k$). It follows that the problem of determining if $g$ is true in every model of $\mathcal{M}$ is in $\Pi_2^P$.

For part (b), let $\text{QBF}_2$ consist of all Quantified Boolean Formulas (QBFs) of the form

$$\forall Q_1 \ldots \forall Q_m \exists P_1 \ldots \exists P_n \varphi,$$

where $\varphi$ is quantifier-free. It is well known that the problem of checking whether a formula in $\text{QBF}_2$ is true is $\Pi_2^P$-complete [Stockmeyer 1977]. We now show how to reduce this problem to the validity problem for $\mathcal{L}_2$.

Let $q = \forall Q_1 \ldots \forall Q_m \exists P_1 \ldots \exists P_n \varphi$ be an arbitrary formula in $\text{QBF}_2$. Let $\varphi'$ be $\varphi$ with $Q_j$ replaced by the ground literal $Q_j(c)$ and $P_k$ replaced by the literal $P_k(x_k)$, for $j = 1, \ldots, m$ and $k = 1, \ldots, n$. It is not hard to see that $q$ is true iff

$$q' = P_1(c) \wedge \ldots \wedge P_n(c) \wedge \neg P_1(c') \wedge \ldots \wedge \neg P_n(c') \Rightarrow \exists x_1 \ldots \exists x_n \varphi'$$

is valid. This follows from two observations. First, note that $q$ is true iff, for every assignment of values to $Q_1, \ldots, Q_m$, there is an assignment of truth values to $P_1, \ldots, P_n$ such that $\varphi$ is true. Second, $q'$ is valid iff, for every interpretation of $Q_1(c), \ldots, Q_m(c)$, there is an assignment of domain elements to $x_1, \ldots, x_n$ such that $\varphi'$ is true. The antecedent $P_1(c) \wedge \ldots \wedge P_n(c) \wedge \neg P_1(c') \wedge \ldots \wedge \neg P_n(c')$ of $q'$ ensures that assigning $x_i$ to $c$ makes $P_i(x_i)$ true, while assigning $x_i$ to $c'$ makes $P(x_i)$ false. Thus, the assignment of values to the variables $x_1, \ldots, x_n$ acts essentially like a truth assignment to $P_1, \ldots, P_n$.

Straightforward manipulations show that $A \Rightarrow B$ is valid iff $A \wedge \neg B \Rightarrow false$ is valid, and $A \wedge \neg B \Rightarrow false$ is valid iff $((\neg A \Rightarrow C) \wedge \neg B) \Rightarrow C$ is valid, provided that none of the nonlogical symbols in $C$ appear in $A$ or $B$. Taking $A$ to be $P_1(c) \wedge \ldots P_n(c) \wedge \neg P_1(c') \wedge \ldots \wedge \neg P_n(c')$, $B$ to be $\exists x_1 \ldots \exists x_n \varphi'$, and $C$ to be $\neg Permitted(d, d')$, where $d$ and $d'$ are distinct from $c$ and $c'$, it follows that $q$ is true iff

$$\forall x_1 \ldots \forall x_n \neg \varphi' \wedge (\neg A \Rightarrow \mathbf{Permitted}(d, d')) \Rightarrow \mathbf{Permitted}(d, d') \qquad (2)$$

is valid. The formula $\neg A \Rightarrow \mathbf{Permitted}(d, d')$ is equivalent to

$$\bigwedge_{i=1}^{n} \neg P_1(c) \Rightarrow \mathbf{Permitted}(d, d') \wedge \bigwedge_{i=1}^{n} P_1(c') \Rightarrow \mathbf{Permitted}(d, d')).$$

Replacing $\neg A \Rightarrow \mathbf{Permitted}(d, d')$ in (2) by the latter formula gives us a formula in $\mathcal{L}_2$. Thus, we have reduced the truth of a QBF formula to the validity of a formula in $\mathcal{L}_2$, as desired.

For part (c), we prove the NP hardness result by reducing the Hamiltonian path problem to the validity problem for $\mathcal{L}_4$. Let $G$ be an undirected graph, where $V = \{v_1, \ldots, v_n\}$ is the set of nodes and $E$ is the set of edges. Let $\Phi$ be a vocabulary that includes the constants $v_1, \ldots, v_n$, a binary predicate $\mathbf{Edge}$, and $\mathbf{Permitted}$. Finally, let $E = \bigwedge_{(v_i, v_j) \in E} \mathbf{Edge}(v_i, v_j)$, and let

$$p = \forall x_1 \ldots \forall x_n (\bigwedge_{i,j \leq n; i \neq j} (x_i \neq x_j) \wedge \bigwedge_{i < n} \mathbf{Edge}(x_i, x_{i+1})) \Rightarrow \mathbf{Permitted}(c, c').$$

It is not hard to show that $E \wedge p \Rightarrow \mathbf{Permitted}(c, c')$ is valid iff there is a Hamiltonian path in $G$. The key observations are (1) there is a Hamiltonian path iff there is an assignment of distinct domain elements to $x_1, ..., x_n$ such that there is an edge between $x_i$ and $x_{i+1}$ for $i < n$, and (2) there is such an assignment iff $E \wedge p \Rightarrow \mathbf{Permitted}(c, c')$ is valid.

We prove the co-NP hardness result by reducing the validity problem for propositional logic to the validity problem for $\mathcal{L}_4$. Let $g$ be a propositional formula, let $v_1, \ldots, v_n$ be the propositions in $g$, and let $g'$ be the first-order formula obtained by replacing the proposition $v_i$ in $g$ with the ground literal $R(c_i)$ for $i = 1, \ldots, n$. It is easy to see that $g$ is valid iff $g'$ is valid. Because $g'$ does not include $\mathbf{Permitted}$, $g'$ is valid iff $g' \vee \mathbf{Permitted}(c, c')$ is valid. Standard manipulations show that $g' \lor \textbf{Permitted}(c, c')$ is equivalent to the $\mathcal{L}_4$ formula $(g' \Rightarrow \textbf{Permitted}(c, c')) \Rightarrow \textbf{Permitted}(c, c')$.   □

## B.   PROOFS FOR SECTION 4

To prove the theorems in Section 4, we need to extend resolution slightly using techniques of *paramodulation* [Robinson and Wos 1983]. Note that if $c$ is the clause $\forall x_1 \dots x_n(c' \lor t_c = t'_c)$, $d$ is the clause $\forall y_1 \dots \forall y_m d'$, $t_d$ is a term in $d'$, and $\sigma$ is a substitution such that $\sigma(t_c) = \sigma(t_d)$, then the following formula is valid:

$$c \land d \Rightarrow \forall x_1 \dots \forall x_n \forall y_1 \dots \forall y_m (c' \lor d'[t_d/t'_c])\sigma. \tag{3}$$

A set of clauses is said to be *closed under paramodulation* if it contains the right-hand side of (3) whenever it contains the clauses on the left-hand side. Let $R^P(f)$ be the set of clauses obtained by closing $f$ under resolution and paramodulation. In other words, $R^P(f)$ is the smallest set of clauses that includes the conjuncts of $f$ (when $f$ is in CNF) and, if we can infer a clause $d$ from two clauses $c$ and $c'$ in $R^P(f)$ by using either resolution or paramodulation, then $d$ is in $R^P(f)$.

THEOREM B.1. [Brand 1975] *If $f$ is a formula in CNF one of whose conjuncts is $\forall x(x = x)$, then $f$ is satisfiable if and only if $R^P(f)$ does not include false.*

We remark that the clause $\forall(x = x)$ is needed here, although it is valid. For example, it is easy to check that $R^P(\forall x(x \neq x))$ does not include *false*, even though $\forall x(x \neq x)$ is not satisfiable. On the other hand, $R^P(\forall(x \neq x) \land \forall(x = x))$ clearly includes *false*.

COROLLARY B.2. *Let $f$ be a CNF formula, none of whose clauses mentions a disjunct of the form $t = t'$. Then $f$ is satisfiable iff $R(f \land \forall x(x = x))$ does not include false.*

PROOF. Clearly $f$ is satisfiable iff $f \land \forall x(x = x)$ is satisfiable. Let $g = f \land \forall x(x = x)$. By Theorem B.1, it suffices to show that $R^P(g) = R(g)$. Clearly, $R(g) \subseteq R^P(g)$. To show that $R^P(g) \subseteq R(g)$, it suffices to show that $R(g)$ is closed under paramodulation. It is not hard to see that, because no clause in $f$ mentions a disjunct of the form $t = t'$, no clause in $R(g) - \{\forall x(x = x)\}$ mentions a disjunct of the form $t = t'$. Therefore, applying paramodulation does not lead to any new clauses.   □

The next four lemmas relate the closures of various formulas and give bounds on the complexity of computing the closure. In these proofs, it is convenient to associate a clause $c$ with its set of disjuncts, which we denote as $\mathcal{S}(c)$. For example, if $\ell_1, \dots, \ell_k$ are literals, then $\mathcal{S}(\ell_1 \lor \dots \lor \ell_k) = \{\ell_1, \dots, \ell_k\}$. For the next four lemmas, let $S = \{s \neq s \mid s \text{ is a term}\}$.

LEMMA B.3. *Let $c$ be a clause with no bipolar literals and let $f$ be a conjunction of ground literals. If a clause $c'$ is in $R(c \land f \land \forall x(x = x))$, then $c'$ is in $R(f \land \forall x(x = x))$ or $\mathcal{S}(c') \subseteq \mathcal{S}(c\sigma) \subseteq \mathcal{S}(c') \cup \mathcal{S}(\neg f) \cup S$ for some substitution $\sigma$.*

PROOF. Let $R'(c \land f)$ consist of the clauses in $R(f \land \forall x(x = x))$ and all clauses $c'$ such that, for some substitution $\sigma$, $\mathcal{S}(c') \subseteq \mathcal{S}(c\sigma) \subseteq \mathcal{S}(c') \cup \mathcal{S}(\neg f) \cup S$. We want to show that $R(c \land f \land \forall x(x = x)) \subseteq R'(c \land f \land \forall x(x = x))$. Because every conjunct of $c \wedge f \wedge \forall x(x = x)$ is in $R'(c \wedge f \wedge \forall x(x = x))$, it suffices to show that $R'(c \wedge f \wedge \forall x(x = x))$ is closed under resolution. To do this, suppose that $c_1$ and $c_2$ are clauses in $R'(c \wedge f \wedge \forall x(x = x))$ that resolve on a literal $\ell$ to create the resolvent $c_3$. We want to show that $c_3 \in R'(c \wedge f \wedge \forall x(x = x))$. If both $c_1$ and $c_2$ are in $R(f \wedge \forall x(x = x))$, then $c_3$ is in $R(f \wedge \forall x(x = x))$, so $c_3 \in R'(c \wedge f \wedge \forall x(x = x))$. If exactly one of the clauses is in $R(f \wedge \forall x(x = x))$, then assume without loss of generality that it is $c_1$. Because $f \wedge \forall x(x = x)$ is a conjunction of literals, every clause in $R(f \wedge \forall x(x = x))$ is either a conjunct of $f \wedge \forall x(x = x)$ or $false$; $c_1$ is the parent of a resolvent, so it is a conjunct of $f \wedge \forall x(x = x)$. Since $c_2 \in R'(c \wedge f \wedge \forall x(x = x)) - R(f \wedge \forall x(x = x))$, there is a substitution $\sigma$ such that $\mathcal{S}(c_2) \subseteq \mathcal{S}(c\sigma) \subseteq \mathcal{S}(c_2) \cup \mathcal{S}(\neg f) \cup S$. Since $c_1$ and $c_2$ are the parents of the resolvent $c_3$ and $c_1$ is a conjunct of $f \wedge \forall x(x = x)$, there is a substitution $\sigma'$ such that $c_2\sigma'$ is $c_3 \vee \sim c_1$, where $\sim c_1$ is the negation of a conjunct of $f$ or has the form $s \neq s$. Because $\mathcal{S}(c_2) \subseteq \mathcal{S}(c\sigma)$, it follows that $\mathcal{S}(c_3) \subseteq \mathcal{S}(c\sigma\sigma')$. Moreover,

$$\mathcal{S}(c\sigma\sigma') \subseteq \mathcal{S}(c_2\sigma') \cup \mathcal{S}(\neg f \sigma') \cup S = \mathcal{S}(c_3) \cup \{\sim c_1\} \cup \mathcal{S}(\neg f) \cup S = \mathcal{S}(c_3) \cup \mathcal{S}(\neg f) \cup S,$$

since $\{(s \neq s)\sigma'|s$ is a term$\} \subseteq S$, $\neg f \sigma' = \neg f$ (because $f$ mentions no variables), and $\sim c_1$ is either a conjunct of $\neg f$ or a literal in $S$. So $c_3 \in R'(c \wedge f \wedge \forall x(x = x))$. Finally, if neither $c_1$ nor $c_2$ is in $R(f \wedge \forall x(x = x))$, then it is not hard to see that there are substitutions $\sigma$ and $\sigma'$ such that $\ell$ is a disjunct of $c\sigma$ and $\neg \ell$ is a disjunct of $c\sigma'$, contradicting the assumption that $c$ has no bipolar literals.  □

For the next three lemmas, let $f = E_0 \wedge \forall x(x = x) \wedge \neg \mathbf{Permitted}(t, t')$ and let $f' = E_1 \wedge P$, where $t$ and $t'$ are closed terms.

LEMMA B.4. $R(f' \wedge f) = \bigcup_{c \in R(f')} R(c \wedge f)$.

PROOF. Let $c$ be a clause in $R(f')$. Because every conjunct of $c \wedge f$ is in $R(f' \wedge f)$ and $R(f' \wedge f)$ is closed under resolution, $R(c \wedge f) \subseteq R(f' \wedge f)$. It follows that $\bigcup_{c \in R(f')} R(c \wedge f) \subseteq R(f' \wedge f)$.

For the opposite inclusion, note that every conjunct of $R(f' \wedge f)$ is in $\bigcup_{c \in R(f')} R(c \wedge f)$. So, it suffices to show that $\bigcup_{c \in R(f')} R(c \wedge f)$ is closed under resolution. To do this, suppose that $c_1, c_2 \in R(f')$ and that $e$ is a resolvent with parents $d_1 \in R(c_1 \wedge f)$ and $d_2 \in R(c_2 \wedge f)$. It suffices to show that $e \in \bigcup_{c \in R(f')} R(c \wedge f)$. If $d_1 \in R(f)$, then clearly $d_1 \in R(c_2 \wedge f)$, so $e \in R(c_2 \wedge f)$ and we are done. Similarly, if $d_2 \in R(f)$, then $e \in R(c_1 \wedge f)$. Suppose that neither $d_1$ nor $d_2$ is in $R(f)$. Then it follows from Lemma B.3 that there are substitutions $\sigma_1$ and $\sigma_2$ such that $c_1\sigma_1 = d_1 \vee d_1'$ and $c_2\sigma_2 = d_2 \vee d_2'$, where $\mathcal{S}(d_1') \subseteq \mathcal{S}(\neg f) \cup S$ and $\mathcal{S}(d_2') \subseteq \mathcal{S}(\neg f) \cup S$. Because $d_1$ and $d_2$ are the parents of $e$, there are substitutions $\sigma_1'$ and $\sigma_2'$, clauses $d_1''$ and $d_2''$, and a literal $\ell$ such that $e = d_1'' \vee d_2''$, $d_1\sigma_1' = d_1'' \vee \ell$, and $d_2\sigma_2' = d_2'' \vee \neg \ell$. Putting the pieces together,

$$c_1\sigma_1\sigma_1' = d_1'' \vee \ell \vee d_1'\sigma_1' \text{ and } c_2\sigma_2\sigma_2' = d_2'' \vee \neg \ell \vee d_2'\sigma_2'.$$

(Note that $\mathcal{S}(d_1'\sigma_1') \subseteq \mathcal{S}(\neg f) \cup S$ and $\mathcal{S}(d_2'\sigma_2') \subseteq \mathcal{S}(\neg f) \cup S$, because the only variables that appear in $d_1'$ or $d_2'$ are in disjuncts of the form $t \neq t$.) Clearly $c_1$ and $c_2$ resolve to create a resolvent $e' \in R(f')$. Moreover, $e'\sigma_1\sigma_2 = e \vee d_1'\sigma_1 \vee d_2'\sigma_2$, so $e \in R(e' \wedge f)$.  □

LEMMA B.5. *If every clause in $f'$ has at most one literal that is bipolar in $f'$, then $R(f')$ has $O(|f'|^2)$ clauses, each of length at most $2L_{f'}L'_{f'}$, and $R(f')$ can computed in time $O(|f'|^2)$.*

PROOF. Note that the resolvent $e$ of two clauses in $f'$ has no bipolars, because every clause in $f'$ has at most one bipolar. It follows that $e$ is not a parent of a resolvent in $R(f')$. So,

$$R(f') = \{c \mid c \text{ is in } \mathcal{S}(f') \text{ or is the resolvent of two clauses in } \mathcal{S}(f')\}.$$

Thus, $R(f')$ has $O(|f'|^2)$ clauses and each clause has length less than $2L_{f'}L'_{f'}$. To find $R(f')$, we simply check each pair of clauses $c$ and $c'$ in $f'$ to see if there is a literal on which they resolve; if so, we resolve them. The check can be done in time $O(|c||c'|)$; the resolution can be done in time $O(|c| + |c'|)$. Since, by assumption, each clause contains at most one instance of a bipolar literal, there will be at most one resolvent for each pair of clauses. It easily follows that $R(f')$ can be computed in time $O(|f'|^2)$. □

If $C$ is a set of clauses, let $\|C\| = \sum_{c \in C} |c|$. For all predicate symbols **Q**, a variable $v$ is **Q**-*constrained* in a clause $c$ if $v$ appears as an argument to **Q** in $c$. Note that a constrained variable, as defined in Section 4.1, is **Permitted**-constrained.

LEMMA B.6. *Suppose that $f$ mentions $m$ terms and $C$ is a non-empty set of clauses such that, for every $c \in C$, no literal in $c$ is bipolar in $c$. Then*

(a) *false $\in \bigcup_{c \in C} R(c \wedge f)$ iff (i) false $\in R(f)$ or (ii) there is a clause $c \in C$ and a substitution $\sigma$ such that $\mathcal{S}(c\sigma) \subseteq \mathcal{S}(\neg f) \cup S$;*

(b) *we can determine whether (i) holds in time $O(|E_0| \log |E_0|)$;*

(c) *we can determine whether (ii) holds in time $O((|E_0| + R\|C\|\|\textbf{Permitted}(t, t')|) \log |E_0|)$, where $R = m$ if every literal in every clause $c$ in $C$ mentions at most one variable that is not constrained in $c$; otherwise $R = m^k$, where every clause $c$ in $C$ has at most $k$ variables that are not constrained in $c$.*

PROOF. For part (a), the "only if" direction follows immediately from Lemma B.3. For the "if" direction, it is easy to see that $R(f) \subseteq R(c \wedge f)$ for every clause $c \in C$. So, if *false* $\in R(f)$, then *false* $\in \bigcup_{c \in C} R(c \wedge f)$. Also, if there is a clause $c \in C$ and a substitution $\sigma$ such that $\mathcal{S}(c\sigma) \subseteq \mathcal{S}(\neg f) \cup S$, then it readily follows from the definition of resolution that *false* $\in R(c \wedge f)$, so *false* $\in \bigcup_{c \in C} R(c \wedge f)$.

For part (b), because $f$ is a conjunction of literals, it is easy to see that *false* $\in R(f)$ iff (1) $E_0$ includes a literal of the form $t \neq t$ or (2) $E_0$ includes a literal and its negation. Clearly, we can check whether (1) holds in time $O(|E_0|)$. To check whether (2) holds, we use a splay tree [Sleator and Tarjan 1983], a form of binary search tree for which, starting with an empty tree, $K$ insertions and $S$ searches take time $O((K + S) \log K)$. Specifically, we insert every negative literal in $E_0$ into the empty splay tree $T$. Then, for every positive literal $\ell$ in $E_0$, we search $T$ for $\neg \ell$. Since at most $|E_0|$ insertion and $|E_0|$ search operations are involved, time $O(|E_0| \log |E_0|)$ is required.

For part (c), recall that $f = E_0 \wedge \forall x (x = x) \wedge \neg\textbf{Permitted}(t, t')$. For any clause $c$, let $c_E$ and $c_P$ be clauses such that $c = c_E \vee c_P$, $c_E$ is **Permitted**-free, and every disjunct in $c_P$ mentions **Permitted**. Because $E_0$ is **Permitted**-free, $\mathcal{S}(c\sigma) \subseteq \mathcal{S}(\neg f) \cup S$ iff $\mathcal{S}(c_E\sigma) \subseteq \mathcal{S}(\neg E_0) \cup S$ and $\mathcal{S}(c_P\sigma) \subseteq \{\mathbf{Permitted}(t,t')\}$. It follows that we can find a substitution $\sigma$ such that $\mathcal{S}(c\sigma) \subseteq \mathcal{S}(\neg f) \cup S$, if one exists, by finding substitutions $\sigma'$ and $\sigma''$ such that $\mathcal{S}(c_P\sigma') \subseteq \{\mathbf{Permitted}(t,t')\}$ and $\mathcal{S}(c_E\sigma'\sigma'') \subseteq \mathcal{S}(\neg E_0) \cup S$, and taking $\sigma = \sigma' \circ \sigma''$ We can assume without loss of generality that $\sigma(x) = x$ for every variable $x$ that does not appear in $c$. Thus, we can clearly check if an appropriate substiuttion $\sigma$ exists in time $O(|c|\|\mathbf{Permitted}(t,t')|)$, by pattern-matching each occurrence of $\mathbf{Permitted}$ in $c_P$ with $\mathbf{Permitted}(t,t')$. Moreover, if $\sigma$ exists, then $|c\sigma| \leq |c|\|\mathbf{Permitted}(t,t')|$, since $\sigma$ substitutes terms in $\mathbf{Permitted}(t,t')$ for variables in $c_P$.

Let

> $D = \{d : \text{there is a clause } c \in C \text{ and a substitution } \sigma \text{ such that } \sigma(x) = x \text{ if } x$
> $\text{does not appear in } c, \mathcal{S}(c_P\sigma) \subseteq \{\mathbf{Permitted}(t,t')\}, \text{ and } d = c_E\sigma\}.$

We can clearly construct $D$ in time $O(\|C\|\|\mathbf{Permitted}(t,t')|)$, by considering the clauses in $C$ one at a time, and $\|D\| < \|C\|\|\mathbf{Permitted}(t,t')|$. Thus, to complete the proof of part (c), it suffices to show that we can determine whether there is a $d \in D$ and a substitution $\sigma$ such that $\mathcal{S}(d\sigma) \subseteq \mathcal{S}(\neg E_0) \cup S$ in time $O((|E_0| + R\|D\|)\log |E_0|)$, where $R$ is as defined in the lemma. We can do this by a brute-force search. In more detail, we insert every literal in $E_0$ into an empty splay tree $T$; then, for each clause $d \in D$ and each possible assignment $\sigma$ of terms in $E_0$ to variables in $d$, we check whether every literal in $d\sigma$ is the negation of a literal in $T$ or is of the form $t \neq t$. Suppose every clause $c$ in $C$ has at most $k$ variables that are not constrained in $c$. Then $d$ has at most $k$ variables. Since $E_0$ mentions at most $m$ terms, it follows that there are at most $m^k$ ways of assigning terms in $E_0$ to variables in $d$. As we have observed, the $O(|E_0|)$ insertions and $O(m^k|d|)$ searches can be done in time $O((|E_0| + m^k|d|)\log |E_0|)$. For each literal $\ell$ in $\mathcal{S}(d\sigma) - \mathcal{S}(\neg E)$, we can determine whether $\ell$ is of the form $t = t$ in time $O(|\ell|)$. Thus, the time needed to check every clause $d \in D$ is $O((|E_0| + m^k\|D\|)\log |E_0|)$.

We may be able to do better if every literal in every clause $c$ in $C$ has at most one variable that is not constrained in $c$. In this case, every literal in every clause $d$ in $D$ has at most one variable. It follows that, given a clause $d \in D$, we can partition the literals in $d$ into sets according to their variable. That is, in time $O(|d|)$, we can write $d$ as $d_1 \vee \ldots \vee d_k$, where two literals $\ell$ and $\ell'$ mention the same variable iff $\ell$ and $\ell'$ both appear in $d_i$ for $i = 1, \ldots, k$. Clearly, there is a substitution $\sigma$ such that $\mathcal{S}(d\sigma) \subseteq \mathcal{S}(\neg E_0) \cup S$ iff there are substitutions $\sigma_1, \ldots, \sigma_k$ such that $\mathcal{S}(d_i\sigma_i) \subseteq \mathcal{S}(\neg E_0) \cup S$, for $i = 1, \ldots, k$. For a particular $d_i$, there are at most $m$ possible substitutions of terms in $E_0$ to the variable in $d_i$. So, given a splay tree $T$ whose entries are the conjuncts in $E_0$, we can determine if there is an appropriate $\sigma_i$ in time $O(m|d_i|\log |E_0|)$. Thus, given $T$, we can determine if there are appropriate substitutions $\sigma_1, \ldots, \sigma_k$ in time $O(m|d|\log |E_0|)$. Since we can construct $T$ in time $O(|E_0|\log |E_0|)$, the total time needed for a particular clause $d \in D$ is $O((|E_0| + m|d|)\log |E_0|)$, and the time needed to check every $d \in D$ is $O((|E_0| + m\|D\|)\log |E_0|)$. $\square$

**PROPOSITION 4.2.** *Suppose that $E \wedge P \Rightarrow \mathbf{Permitted}(t,t')$ is a standard query in which $E$ is basic, the equality symbol is not mentioned in $E \wedge P$, and there are no bipolars in $P$. Then $E \wedge P \Rightarrow \mathbf{Permitted}(t,t')$ is valid iff there is a conjunct $p$* *of $P$ such that $E \wedge p \Rightarrow \mathbf{Permitted}(t, t')$ is valid.*

Proof. Here and elsewhere, let $E^+$ be an abbreviation for $\forall x (x = x) \wedge E$. By Corollary B.2, it suffices to show that $R(E^+ \wedge P \wedge \neg\mathbf{Permitted}(t, t'))$ includes *false* iff $R(E^+ \wedge p \wedge \neg\mathbf{Permitted}(t, t'))$ includes *false* for some conjunct $p$ of $P$. It follows from Lemma B.4, where we take $f$ to be $E^+ \wedge \neg\mathbf{Permitted}(t, t')$ and $f'$ to be $P$, that $R(E^+ \wedge P \wedge \neg\mathbf{Permitted}(t, t'))$ includes *false* iff $R(E^+ \wedge c \wedge \neg\mathbf{Permitted}(t, t'))$ includes *false* for some $c$ in $R(P)$. Since there are no bipolar literals in $P$, $R(P)$ is just the set of conjuncts in $P$, so we are done.  □

Theorem 4.1. *Let $\mathcal{L}_5$ consist of all standard queries of the form $E \wedge P \Rightarrow \mathbf{Permitted}(t, t')$ such that*

*(1) $E$ is basic (i.e., $E$ is a conjunction of ground literals),*

*(2) there are no bipolar literals in $P$,*

*(3) equality is not mentioned in $E \wedge P$, and*

*(4) every variable appearing in a conjunct $p$ of $P$ is constrained in $p$.*

*We can determine the validity of formulas in $\mathcal{L}_5$ in time $O((|E| + |P||\mathbf{Permitted}(t, t')|) \log |E|)$, where $|\varphi|$ denotes the length of $\varphi$, when viewed as a string of symbols.*

Proof. Let $S_p$ be the set of conjuncts of $P$. By Proposition 4.2, $E \wedge P \Rightarrow \mathbf{Permitted}(t, t')$ is valid iff $E \wedge p \Rightarrow \mathbf{Permitted}(t, t')$ is valid, for some conjunct $p$ of $P$. By Corollary B.2, the latter statement holds iff *false* $\in \bigcup_{p \in S_p} R(E^+ \wedge p \wedge \neg\mathbf{Permitted}(t, t'))$. It follows from Lemma B.6(a), where we take $C = S_p$ and $f = E^+ \wedge \neg\mathbf{Permitted}(t, t')$, that *false* $\in \bigcup_{p \in S_p} R(E^+ \wedge p \wedge \neg\mathbf{Permitted}(t, t'))$ iff (a) *false* is in $R(E^+ \wedge \neg\mathbf{Permitted}(t, t'))$ or (b) there is a clause $p \in S_p$ and a substitution $\sigma$ such that $\mathcal{S}(p\sigma) \subseteq \mathcal{S}(\neg E^+ \vee \mathbf{Permitted}(t, t')) \cup \{s \neq s \mid s \text{ is a term}\}$. By Lemma B.6(b), we can determine whether (a) holds in time $O(|E| \log |E|)$. It follows from Lemma B.6(c), where $f = E^+ \wedge \neg\mathbf{Permitted}(t, t')$, $C = S_p$, and $k = 0$, that we can determine whether (b) holds in time $O((|E| + |P||\mathbf{Permitted}(t, t')|) \log |E|)$.  □

Rather than just proving Theorem 4.5, we prove a slightly stronger result, from which we will also be able to prove Theorem 5.1. Note that part (b) of the following theorem is equivalent to Theorem 4.5.

Theorem B.7. *Suppose that $E$ is a standard environment, $P$ is a conjunction of pure permitting policies, and $D$ is a conjunction of (not necessarily pure) denying policies such that, for every resolvent $f$ created by resolving a conjunct of $P$ and a conjunct of $D$ on a literal that mentions $\mathbf{Permitted}$, either $E \Rightarrow f$ is valid or $q \Rightarrow f$ is valid for some conjunct $q$ of $P \wedge D$. Then*

*(a) $E \wedge P$ is consistent iff $E \wedge P \wedge D$ is consistent*

*(b) $E \wedge P \wedge \neg\mathbf{Permitted}(t, t')$ is consistent iff $E \wedge P \wedge D \wedge \neg\mathbf{Permitted}(t, t')$ is consistent, where $t$ and $t'$ are terms of the appropriate sort.*

Proof. We prove part (a) here; the proof of part (b) is identical.

Suppose that the hypotheses of the theorem hold. Since $g = E \wedge \bigwedge_{p \in P} p$ is consistent, it has a *Herbrand model*, that is, a model whose domain consists of all the variable-free terms in the language. Of the Herbrand models for $g$, let $m$ be a *minimally permissive* one, that is, one for which the extension of the **Permitted** predicate is minimal. We claim that in fact $m \models E \wedge \bigwedge_{p \in P} p \wedge \bigwedge_{d \in D} d$. For suppose not. Then there is a denying policy $\forall x_1 \ldots \forall x_n d$ in $D$ and a variable substitution $\sigma_d$ such that:

(1) $m \models \neg d\sigma_d$ and

(2) for all denying policies $\forall x_1 \ldots \forall x_n e \in D$ and variable substitution $\sigma_e$ such that $m \models \neg e\sigma_e$, the number of negative literals in $e\sigma_e$ that mention **Permitted** is at least the number of negative literals in $d\sigma_d$ that mention **Permitted**. (Note that we are assuming all policies are in CNF, so that the number of negative literals in a policy is well-defined.)

Since $d$ is a denying policy, $d\sigma_d$ has at least one negated **Permitted** formula among its clauses; that is $d\sigma_d = d' \vee \neg\textbf{Permitted}(s_d, s'_d)$ for some terms $s_d$ and $s'_d$. Since $m \models \neg d\sigma_d$, we have that $m \models \neg d' \wedge \textbf{Permitted}(s_d, s'_d)$. Since $m$ is minimally permissive, there must be a pure permitting policy $\forall x_1 \ldots \forall x_n p \in P$ and a variable substitution $\sigma_p$ such that $p\sigma_p = p' \vee \textbf{Permitted}(s_d, s'_d)$ and $m \models \neg p'$. (Otherwise, consider the model obtained by removing $(s_d, s'_d)$ from the extension of **Permitted**; it must also satisfy $g$, and is less permissive than $m$.) Let $f = p' \vee d'$ be the formula created by resolving $p\sigma_p$ and $d\sigma_d$ on **Permitted**$(s_d, s'_d)$. Note that, by choice of $p'$ and $d'$, $m \models \neg f$. It follows from the definition of resolution and the fact that $p$ is a *pure* permitting policy that the number of negative literals in $f$ that mention **Permitted** is less than the number of such literals in $d\sigma_d$. Moreover, by hypothesis, either $E \Rightarrow f$ is valid or $q \Rightarrow f$ is valid for some $q \in P \cup D$. Since $m \models E \wedge \neg f$, $E \Rightarrow f$ is not valid, so $q \Rightarrow f$ is valid for some $q \in P \cup D$. Since $m \models \neg f$, $m \models \neg q$; and, since $m \models \bigwedge_{p \in P} p$ by assumption, $q \in D$. Therefore, there is a denying policy $\forall x_1 \ldots \forall x_n e \in D$ such that $\forall x_1 \ldots \forall x_n e \Rightarrow f$ is valid. It is not hard to show that $\forall x_1 \ldots \forall x_n e \Rightarrow f$ is valid iff there is a variable substitution $\sigma_e$ such that $e\sigma_e = f$. Thus, $\forall x_1 \ldots \forall x_n e$ is a denying policy in $D$ and $\sigma_e$ is a variable substitution such that $m \models \neg e\sigma_e$ (because $m \models \neg f$). The number of negative literals in $e\sigma_e$ that mention **Permitted** (which is the number of negative literals in $f$ that mention **Permitted**) is less than the number of such literals in $d\sigma_d$. Thus, we have a contradiction. □

PROPOSITION 4.6. *If $q$ is an equality-safe standard query, then there is a standard query $q'$ of the form $E'_0 \wedge E'_1 \wedge P' \Rightarrow \textbf{Permitted}(t, t')$ such that (a) $q$ is valid iff $q'$ is valid, (b) $q'$ is equation-free, and (c) $|q'| = O(|q||L'_q|)$, where $L'_q$ is the length of the longest term in $q$. Moreover, we can find such a $q'$ in time $O(|q|)$.*

PROOF. Suppose that $q$ has the form $F_0 \wedge F_1 \wedge E_1 \wedge P \Rightarrow \textbf{Permitted}(s, s')$, where $F_0$ is the conjunction of the equality statements, while $F_1$ consists of the remaining conjuncts in $E_0$. To create $q'$, we partition the set of terms in $E_0$ into equivalence classes; terms $t_e$ and $t'_e$ are in the same class if the equality formulas in $E_0$ imply $t_e = t'_e$. The equivalence classes can be found in linear time.[4] Since $q$ is equality-safe, each equivalence class has at most one term that is not a constant. If an equivalence class has a term that is not a constant, then we choose that term to represent the class; otherwise, we select a representative arbitrarily. Let $q' = F_1' \wedge E_1' \wedge P' \Rightarrow \mathbf{Permitted}(t, t')$, where $F_1'$, $E_1'$, $P'$, $t$, and $t'$ are the result of replacing each closed term in $F_1$, $E_1$, $P$, $s$, and $s'$, respectively, that also appears in $F_0$ by its representative. It suffices to show that $q$ is valid if and only if $q'$ is valid, because the other statements in the conclusion of Proposition 4.6 follow immediately from the construction of $q'$. Let $q'' = F_0 \wedge F_1' \wedge E_1' \wedge P' \Rightarrow \mathbf{Permitted}(t, t')$. It is easy to see that $q$ is equivalent to $q''$; substituting a term by its representative in the equivalence class is justified in the presence of $F_0$. Thus, it suffices to show that $q''$ is valid iff $q'$ is valid. The "if" direction is trivial. For the "only if" direction, suppose by way of contradiction that $q''$ is valid and $q'$ is not. It follows that there is a model $m$ with interpretation $I$ that does not satisfy $q'$. Let $m'$ be a model that is identical to $m$ except that $m'$ interprets a constant $r$ as $I(r')$ if $r$ and $r'$ are in the same equivalence class and $r'$ is the class representative. Clearly, $m'$ satisfies $F_0$. Moreover, because $m$ does not satisfy $q'$ and the only difference between $m$ and $m'$ is the interpretation of constants that are not mentioned in $q'$, $m'$ does not satisfy $q'$. This contradicts the validity of $q'$.  □

---

[4] In general, the problem of constructing equivalence classes is harder than linear time. For example, if $c_1 = c_2$, then $f(c_1) = f(c_2)$. However, we do not have to worry about drawing such inferences—if $E_0 \Rightarrow (c_1 = c_2)$ is valid, then it cannot be the case that $f(c_1)$ and $f(c_2)$ are both terms in $E_0$, for then $E_0 \Rightarrow (f(c_1) = f(c_2))$ is valid, and $q$ would not be equality-safe.

The following example illustrates the procedure for creating $q'$ from $q$.

EXAMPLE B.8. *Consider the query "may Bob nap", given that Alice is Bob's wife, Alice may nap, and any individual may nap if his wife may nap. We can write the query as $q = e \wedge p_1 \wedge p_2 \Rightarrow \mathbf{Permitted}(\mathbf{Bob}, \mathbf{nap})$, where*

$$e = (\mathbf{Alice} = \mathbf{wifeOf}(\mathbf{Bob})),$$
$$p_1 = \mathbf{Permitted}(\mathbf{Alice}, \mathbf{nap}), \ and$$
$$p_2 = \forall x(\mathbf{Permitted}(\mathbf{wifeOf}(x), \mathbf{nap}) \Rightarrow \mathbf{Permitted}(x, \mathbf{nap})).$$

*The query $q'$ is the result of removing the conjunct $e$ from $q$ and replacing every occurrence of $\mathbf{Alice}$ by $\mathbf{wifeOf}(\mathbf{Bob})$. Thus, $q' = p_1' \wedge p_2' \Rightarrow \mathbf{Permitted}(\mathbf{Bob}, \mathbf{nap})$, where*

$$p_1' = \mathbf{Permitted}(\mathbf{wifeOf}(\mathbf{Bob}), \mathbf{nap}) \ and$$
$$p_2' = \forall x(\mathbf{Permitted}(\mathbf{wifeOf}(x), \mathbf{nap}) \Rightarrow \mathbf{Permitted}(x, \mathbf{nap})).$$

*Note that we replace $\mathbf{Alice}$ by $\mathbf{wifeOf}(\mathbf{Bob})$ because the two terms are in the same equivalence class and, since $\mathbf{wifeOf}(\mathbf{Bob})$ mentions a function symbol, it is the class representative. Also note that if we replace $\mathbf{wifeOf}(\mathbf{Bob})$ by $\mathbf{Alice}$, then the resulting query is not valid, even though $q$ is. In general, we do not preserve validity if we replace a term that includes a function symbol. That is why we restrict to equality-safe queries in Proposition 4.6.* ∎

THEOREM 4.7. *The validity of an equation-free Lithium query $q = E_0 \wedge E_1 \wedge P \Rightarrow \mathbf{Permitted}(t, t')$ with $m$ terms in $E_0$ can be determined in time $O((|E_0| + T|E_1 \wedge P|^2) \log |E_0|)$, where $T = mL_{E_1 \wedge P} L'_{E_1 \wedge P} |\mathbf{Permitted}(t, t')|$ if every literal in every conjunct $c$ of $E_1 \wedge P$ mentions at most one variable that is not constrained in $c$ relative to $q$; otherwise, $T = m^{2k} L_{E_1 \wedge P} L'_{E_1 \wedge P} |\mathbf{Permitted}(t, t')|$, where every conjunct $c$ of $E_1 \wedge P$ has at most $k$ variables that are not constrained in $c$ relative to $q$.*

Proof. By Corollary B.2, $q$ is valid iff $R(E_0 \wedge \forall x(x = x) \wedge E_1 \wedge P \wedge \neg \textbf{Permitted}(t, t'))$ includes *false*. Let $E_0^+$ be $E_0 \wedge \forall x(x = x)$ and let $q^+$ be the result of replacing $E_0$ in the antecedent of $q$ by $E_0^+$. By Lemma B.4, *false* $\in R(\neg q^+)$ iff there is a clause $c \in R(E_1 \wedge P)$ such that *false* $\in R(c \wedge E_0^+ \wedge \neg \textbf{Permitted}(t, t'))$. By Lemma B.6(a), the latter statement holds iff (1) *false* $\in R(E_0^+ \wedge \neg \textbf{Permitted}(t, t'))$ or (2) there is a clause $c \in R(E_1 \wedge P)$ and a substitution $\sigma$ such that $\mathcal{S}(c\sigma) \subseteq \mathcal{S}(\neg E_0^+ \vee \textbf{Permitted}(t, t')) \cup \{s \neq s \mid s \text{ is a term}\}$. By Lemma B.6(b), we can check whether (1) holds in time $O(|E_0| \log |E_0|)$. To determine whether (2) holds, we first note that, by Lemma B.5, we can compute $R(E_1 \wedge P)$ in time $O(|E_1 \wedge P|^2)$. Once we have $R(E_1 \wedge P)$, it follows from Lemma B.6(c), where we take $C = R(E_1 \wedge P)$, that we can determine whether (2) holds in time $O((|E_0| + m^{k'}|R(E_1 \wedge P)||\textbf{Permitted}(t, t')|) \log |E_0|)$ if every clause $c \in R(E_1 \wedge P)$ has at most $k'$ variables that are not constrained in $c$. It follows from Lemma B.5 that $|R(E_1 \wedge P)|$ is $O(|E_1 \wedge P|^2 L_{E_1 \wedge P} L'_{E_1 \wedge P})$; it follows from the way resolution is defined that $k' \leq 2k$. So, we can determine whether (2) holds in time $O((|E_0| + m^{2k}|E_1 \wedge P|^2 L_{E_1 \wedge P} L'_{E_1 \wedge P}|\textbf{Permitted}(t, t')|) \log |E_0|)$.

Suppose that every literal in every conjunct $c$ of $E_1 \wedge P$ mentions at most one variable that is not constrained in $c$ relative to $q$. Then it follows from the definition of resolution that every literal $\ell$ in every clause $c$ in $R(E_1 \wedge P)$ mentions at most one variable that is not constrained in $c$. It follows from Lemma B.6(c), where we again take $C = R(E_1 \wedge P)$, that we can determine whether (2) holds in this case in time $O((|E_0| + m|R(E_1 \wedge P)||\textbf{Permitted}(t, t')|) \log |E_0|)$, once we have computed $R(E_1 \wedge P)$. By Lemma B.5, we can compute $R(E_1 \wedge P)$ in time $O(|E_1 \wedge P|^2)$ and $|R(E_1 \wedge P)|$ is $O(|E_1 \wedge P|^2 L_{E_1 \wedge P} L'_{E_1 \wedge P})$. So, the total time needed to determine whether (2) holds is $O((|E_0| + m|E_1 \wedge P|^2 L_{E_1 \wedge P} L'_{E_1 \wedge P}|\textbf{Permitted}(t, t')|) \log |E_0|)$. ☐

## C. PROOFS FOR SECTION 5

Theorem 5.1. *Suppose that $E$ is an environment, $P$ is a conjunction of pure permitting policies, and $D$ is a conjunction of (not necessarily pure) denying policies such that the antecedent of Theorem 4.5 holds. Then $E \wedge P \wedge D$ is satisfiable iff $E$ is satisfiable.*

Proof. If $E$ is satisfiable, then $E \wedge P$ satisfiable. (For any model $m$ that satisfies $E$ there is a model $m'$ that is identical to $m$, except $m'$ satisfies $\textbf{Permitted}(s, s')$ for all terms $s$ and $s'$ of the appropriate sort; $m'$ satisfies $E \wedge P$.) The result is now immediate from Theorem B.7(a). ☐

### Acknowledgments

We would like to thank Carl Lagoze for his advice on the policy needs of digital libraries, Riccardo Pucella for numerous discussions on the material presented here, Thomas Bruce for pointing us to the documents on Social Security, Kevin O'Neill for suggesting the name Lithium, John Mitchell and Moshe Vardi for discussions about the complexity of fragments of first-order logic, and Michael Tschantz for reviewing an earlier version of this paper.